\documentclass[11pt,tightenlines,eqsecnum,floats,aps,amsfonts,amsmath,amssymb,nofootinbib,prd,shownopacs,floatfix,noeprint,notitlepage]{revtex4-1}
\usepackage{graphicx}
\usepackage{epstopdf}
\usepackage{latexsym}
\usepackage{amssymb}
\usepackage{amsmath}
\usepackage{color}
\usepackage{mathrsfs}
\usepackage{xparse}
\usepackage{float}
\usepackage{mathtools}
\usepackage{bbold}
\usepackage[section]{placeins}
\usepackage{setspace}
\usepackage[center]{subfigure}

\makeatletter
\def\p@subsection{}
\makeatother
\begin{document}
\newcommand{\ov}{\mathcal{O}_V}
\newcommand{\ob}{\mathcal{O}_b}
\newcommand{\of}{\mathcal{O}_\varphi}
\newcommand{\opf}{\mathcal{O}_{\pi_\varphi}} 
\newcommand{\der}{{\mathrm{d}}}
\newcommand{\lp}{{\ell_{\mathrm{Pl}}}}
\newcommand{\be}{\begin{equation}}
\newcommand{\ee}{\end{equation}}
%%%%%%%%%%%%%%%%%%%%%%%%%%%%%%%%%%%%%%%%%%%%%%%%%%%%%%%%%%%%%%%
\renewcommand\arraystretch{2}
\newcommand{\bq}{\begin{equation}}
\newcommand{\eq}{\end{equation}}
\newcommand{\bqn}{\begin{eqnarray}}
\newcommand{\eqn}{\end{eqnarray}}
\newcommand{\nb}{\nonumber}
\newcommand{\lb}{\label}
\newcommand{\cb}{\color{blue}}
\newcommand{\cc}{\color{cyan}}
\newcommand{\cm}{\color{magenta}}
\newcommand{\rc}{\rho^{\scriptscriptstyle{\mathrm{I}}}_c}
\newcommand{\rd}{\rho^{\scriptscriptstyle{\mathrm{II}}}_c} 
\NewDocumentCommand{\evalat}{sO{\big}mm}{%
  \IfBooleanTF{#1}
   {\mleft. #3 \mright|_{#4}}
   {#3#2|_{#4}}%
}
\newcommand{\PRL}{Phys. Rev. Lett.}
\newcommand{\PL}{Phys. Lett.}
\newcommand{\PR}{Phys. Rev.}
\newcommand{\CQG}{Class. Quantum Grav.}
\newcommand{\parallelsum}{\mathbin{\!/\mkern-5mu/\!}}
 %%%%%%%%%%%%%%%%%%%%%%%%%%%%%%%%%%%%%%%%%%%%%%%%%%%%%%%%%%%%%%%
\title{Towards a reduced phase space quantization in loop quantum cosmology with an inflationary potential} 
\author{Kristina Giesel$^{1}$}
\email{kristina.giesel@gravity.fau.de}
\author{Bao-Fei Li $^{2}$}
\email{baofeili1@lsu.edu}
\author{Parampreet Singh$^2$}
\email{psingh@lsu.edu}
\affiliation{$^{1}$ Institute for Quantum Gravity, Department of Physics, FAU Erlangen-N\"urnberg, Staudtstr. 7, 91058 Erlangen, Germany\\
$^{2}$ Department of Physics and Astronomy, Louisiana State University, Baton Rouge, LA 70803, USA}
%\date{today}
\begin{abstract}
We explore a reduced phase space quantization of loop quantum cosmology (LQC) for a spatially flat FLRW universe filled with reference fields and an inflaton field in a Starobinsky inflationary potential. We consider three separate cases in which the reference fields are taken to be the Gaussian dust, the  Brown-Kucha\v{r}  dust, and massless Klein-Gordon scalar reference fields respectively. 
This is a ``two-fluid" model in which reference fields 
act as global clocks providing a physical time in an inflationary spacetime, and  
allow bypassing various technical hurdles in conventional quantum cosmological models. The reduced phase space is obtained in terms of the Dirac observables of the gravitational as well as the inflaton degrees of freedom. The physical Hamiltonians of the two dust models take the same form but turn out to be quite different from that of the  Klein-Gordon reference field which reflects an aspect of the multiple choice problem of time. Loop quantization is implemented using the so-called $\bar \mu$ scheme and the Schr\"{o}dinger equations involving the physical Hamiltonian operators generating the evolution in the physical time in the dust and massless Klein-Gordon models are obtained. These turn out to be quantum difference equations with same non-singular structure as for other models in LQC. We study some phenomenological implications of the quantization using the effective dynamics resulting from the reduced phase space quantization including the resolution of the big bang singularity via a quantum bounce, and effects of the different reference fields on e-foldings in both the pre-inflationary and the slow-roll inflationary phases. We find that different clocks, even when starting with a small but same energy density, can leave tiny but different imprints on the inflationary dynamics. In addition, for Brown-Kucha\v{r} dust, the choice of a negative energy density can result in a cyclic evolution before the onset of inflation constraining certain values for ideal clocks.
\end{abstract}
\maketitle
\section{Introduction}
\label{Intro}
\renewcommand{\theequation}{1.\arabic{equation}}\setcounter{equation}{0}
The problem of time in canonical quantum gravity is complex and comes in different forms, which include the existence of a global time, and the multiple choice problem  \cite{kuchar1991}. Since these issues affect the way one interprets predictions from quantum gravity and quantum cosmology, they are necessary to be squarely addressed to obtain a consistent dynamics and phenomenology. In the quantization of GR as a totally constrained system, a related  important issue which one encounters is to find the physical Hilbert space equipped with a physical inner product. In general, this is a non-trivial task, such as in the Dirac's method of the quantization of constrained systems where it requires a successful implementation of the refined algebraic quantization or the group averaging procedure \cite{refined}. In certain situations, which are of significant physical interest, all above issues need to be addressed together. One such situation arises in the presence of inflationary potentials where these issues become intricate, such as what is the choice of appropriate clocks, how to satisfy a global time requirement in inflationary dynamics and how it affects unitarity, and how to construct the associated conserved physical inner product. 

To overcome the problem of time in quantum gravity, a strategy is the relational formalism in which reference fields are introduced to extract dynamics of the remaining degrees of freedom  \cite{Vy1994, dittrich, dittrich2,bergman1961,*bergman2,*komar1958,*kuchar1991,*Isham:1992ms,*rovelli,*rovelli2, *Pons:2009cz,*Pons:2010ad, *Anderson:2010xm}. Relevant observables can be constructed from  phase space functions via an observable map introduced in \cite{Vy1994,dittrich}. The resulting  gauge-invariant quantities, so-called Dirac observables Poisson commute with all first class constraints. The relational formalism has been successfully used in various settings to extract dynamics in GR  
 \cite{dittrich2,dt2006,ghtw2010I,ghtw2010II,gt2015,Ali:2015ftw,gh2018,ghs2018,gsw2019,laura}, scalar-tensor theories \cite{hgm2015}, Lema\^itre-Tolman-Bondi spacetimes \cite{gtt2010}, loop quantum gravity (LQG) \cite{gt2007,dgkl2010,hp2011,hp2012,gv2016,gv2017,Ali:2018vmt,hl2019I,hl2019II} and quantum cosmological models (see for eg.  \cite{aps2006a,aps2006b,aps2006c,Kaup,*Blyth,*acs2010,*closed1,*open,*ads,*ds1,*ds2,*radiation,*Mielczarek:2010rq,*Amemiya:2009pj,*Gryb:2018whn}). The latter provide a simpler setting where the Gauss and spatial-diffeomorphisms are fixed in a symmetry reduction, and a  single massless scalar field is most widely used as a reference field both in the Wheeler-DeWitt quantum cosmology \cite{Kaup,Blyth} and LQC \cite{aps2006a,aps2006b,aps2006c,acs2010,*as2011}. In LQC,  the relational dynamics has so far been used in absence of any scalar potential and physical evolution has been explored for various spacetimes \cite{aps2006a,*aps2006b,*aps2006c,*acs2010,*closed1,*open,*ds1,*ds2,*ads,*radiation,*as2011}. Expectation values of Dirac observables, such as volume and energy density, show that the big bang singularity is resolved via a quantum bounce occurring when the spacetime curvature reaches Planckian values. Using consistent histories formulation, probability for bounce to occur turns out to be unity \cite{Craig:2013mga}. Based on these results various extensions have been successfully pursued, including investigations on a generic resolution of singularities  \cite{s2009,s2014}, quantization of  anisotropic models \cite{awe}, inclusion of inhomogeneities in Gowdy models \cite{MartinBenito:2008ej,gowdy}, and various phenomenological implications have been studied (see \cite{as2017} for a review). Despite this remarkable success, it is difficult to adopt above strategy to inflationary spacetimes because of a resulting time-dependent Hamiltonian constraint if one employs a single scalar field as is usually done in LQC, which in this case is the inflaton,  as the clock. This  complicates the task of finding the physical Hilbert space equipped with a conserved inner product in the time-dependent case.\footnote{If the effect of the potential is assumed to be small in the pre-inflationary epoch, one can consider a local time range in which the inflaton clock behaves monotonically and one finds that the big bang singularity is resolved \cite{aps4}. While such an approach does not provide a full quantum gravitational treatment of an  inflationary spacetime where above issues are resolved, it does allow to gain some insights on the physics of bounce if potential plays little role \cite{aps4} and associated phenomenology \cite{aag}, which has proved  useful to get glimpses of  quantum gravity effects via cosmological perturbations (see for eg. \cite{dressed,Fernandez-Mendez:2013jqa,as2017}).} 
Given that the inflationary paradigm is widely considered essential to explain cosmological observations, and since it is past-incomplete in classical GR, obtaining a quantization of inflationary spacetimes which is non-singular is an important open issue. Our work aims to fill this gap in LQC.

For the quantization of inflationary spacetimes in LQC we use an 
alternative strategy to Dirac quantization known as the reduced phase space quantization which bypasses various difficulties of the former. Here one chooses appropriate reference fields and reduces (part of) the constraints already at the classical level by constructing Dirac observables for just the remaining degrees of freedom. The quantization of the resulting reduced phase space yields the physical Hilbert space paralleling situation for unconstrained theories with a physical Hamiltonian operator that does not vanish.  The latter is proportional to the momentum of the clock and is a constant of motion in the FLRW models considered here. In reduced phase space quantization one can either reduce the Hamiltonian and spatial diffeomorphism constraints  or just the Hamiltonian constraint at the classical level using either four or one reference fields respectively. These two models are classified as type I and type II respectively \cite{gt2015}, where for the former physical Hilbert space is obtained directly but for the latter one requires to partially use the Dirac quantization \cite{dgkl2010,*hp2012,*Giesel:2006uj,*jurek3}. Since diffeomorphisms are trivially vanishing in FLRW model, differences between type I \cite{gv2016,gv2017}  and type II models  \cite{dgkl2010,hp2012,hp2011} disappear at the level of the background dynamics. But these differences become important at the level of perturbations (see for eg. \cite{ghls2020}). In our analysis we consider three different type I models with reference fields using the Gaussian dust \cite{Kuchar:1990vy,gt2015}, the Brown-Kucha\v{r} dust \cite{bk1995,ghtw2010I,gt2007} and the four scalar field model \cite{gv2016,gv2017} with an inflaton in Starobinsky inflationary potential which is favored by observations. In the homogeneous and isotropic models, due to symmetry reduction, each of these models contains one reference field and one inflaton field, both of which are minimally coupled to gravity. %correspond to one reference field minimally coupled to inflaton.
Note that unlike earlier works inflaton is not chosen as a reference field (as in \cite{aps4,Fernandez-Mendez:2013jqa}). An advantage of our method is that the resulting physical Hamiltonian is  time-independent, whereas this is no longer the case when the inflaton is chosen as the clock with a non-trivial potential.
%Our analysis  can be extended easily to Wheeler-DeWitt quantum cosmology as well as  other potentials, including those used as alternatives to inflation.
%Models of type I that can be considered in this context are for instance the Gaussian dust \cite{Kuchar:1990vy,gt2015}, the Brown-Kucha\v{r} dust \cite{bk1995,ghtw2010I,gt2007} and the four scalar field model \cite{gv2016,gv2017}, where the latter in this work  has been extended by the coupling of an inflaton field which can be easily included into the constraint analysis. 
Though our work uses  established techniques in LQC, it is based on a reduced phase space quantization unlike the Dirac quantization mainly used so far in LQC.\footnote{A model of type II that also provides a time-independent physical Hamiltonian even if the inflaton potential is considered is the one in \cite{hp2012,hp2011}.  Similar to the four scalar field model \cite{gv2016,gv2017}, the type II model in \cite{dgkl2010}, although not analyzed in this direction so far, could also be generalized to more degrees of freedom including an inflaton coupled in addition to the reference field but for the model in \cite{dgkl2010} this needs to be implemented at the level of the diffeomorphism invariant Hilbert space and it would be interesting to compare such a generalizations to the generalized model of \cite{gv2016,gv2017}} 

The models that will be considered here have the common property that the algebra of the Dirac observables has the standard canonical form, which in general need not be the case. This is an important advantage as far as the reduced phase space quantization is considered because in this case we can use the standard LQC representation, usually used for the kinematical Hilbert space in a Dirac quantization approach, as the representation for the physical Hilbert space here generalized to the case of two-fluid models.  This is in contrast to the Dirac quantization, where obtaining the physical Hilbert space is non-trivial and involves finding an inner product using a group averaging procedure. 
Interestingly, due to the property of the observable map that the observable of a function of the elementary phase space variables is just the function of the corresponding Dirac observables, the reduced phase space quantization of LQC shares many features of the polymer quantization in standard LQC but needs to be generalized to the two fluid case due to the additional degrees of freedom we start with by coupling the reference fields in addition to the inflaton. The advantage of this is that we can formulate the quantum dynamics governed by a Schr\"odinger-like  equation directly in the physical Hilbert space and the physical inner product can easily be constructed. This is in contrast to studies in standard LQC conducted so far, where one uses the inflaton itself as a clock and the understanding of the physical Hilbert space of a time-dependent quantum constraint is rather complicated. Note that the inner product used for models with a massless scalar field is no longer valid for inflationary spacetimes. 

Investigations of inflationary potentials, including the Starobinsky potential have been undertaken earlier in LQC but neither there is any study so far at the full quantum level where the physical Hilbert space equipped with a physical inner product is known, nor there are investigations with clocks in addition to the inflaton. Note that in the conventional setting of Starobinsky inflation one starts from an $R^2$ action and obtains an equivalent scalar field potential in the Einstein frame. This strategy has not yet been replicated in LQC where the higher order action is obtained in a Palatini framework \cite{olmo-ps}. Therefore, as in various other works in LQC, we assume the existence of a Starobinsky scalar field potential in our analysis. While the form of this potential may change when quantum gravitational generalization of above construction is available, this would not affect the validity of our approach.  
%Note that unlike the conventional derivation, the Starobinsky potential is not derived in LQC from a higher curvature term in the action. Rather, as in other works in LQC we assume its  it as a phenomenological input. For a discussion of higher curvature terms in the action framework of LQC, see \cite{olmo-ps}.}. 
Further, the physical Hamiltonian for the Gaussian and Brown-Kucha\v{r} dust models in the classical FLRW spacetime with a Starobinsky potential has been studied recently in \cite{ghls2020}. It turns out that due to the homogeneity and the isotropy of the spacetime, the background evolution of the universe obeys the same equations of motion in both dust models for the classical theory, but, differences arise when linear perturbations around a FLRW spacetime are considered. This is not surprising because Gaussian and Brown-Kucha\v{r} dust clocks have same variation of energy density with respect to the scale factor and any differences in background dynamics vanish unless one considers dust clocks with negative energy density which are permissible in the latter case. In contrast, since the energy density of Klein-Gordon reference field has different variation with respect to scale factor when compared to dust clocks, there can arise small differences in physical predictions for different clocks. Our analysis will also deal with understanding some of these differences in pre-inflation and inflationary epochs.  
%Although introducing the dust reference fields avoids the problem of time in GR, the big bang singularity in the cosmological setting still remains. 

This paper is organized as follows. In section \ref{sec:review}, we give a brief review of the relational formalism using the Gaussian and the  Brown-Kucha\v{r} dust models and the Klein-Gordon scalar field model in GR. The reduced phase space constructed from the observable map in these models is discussed in a concise way and the physical Hamiltonian will be given in each case. In order to analyze these models in the context of cosmology, we  also explicitly discuss the symmetry reduced case of spatially flat FLRW spacetime with the resulting physical Hamiltonians. 
In section \ref{sec:LQC},  the reduced phase space of the flat FLRW universe is formulated in terms of the Dirac observables corresponding to holonomies and the triads and the physical Hamiltonians in the reduced phase space   are expressed in terms of these fundamental Dirac observables. The quantization is then carried out in the $\bar \mu $ scheme used in LQC which  results in non-singular quantum difference  equations formulated at the level of the physical Hilbert space with the same structure as in conventional LQC. In order to facilitate a numerical analysis, we also assume the validity of the effective dynamics of the quantized background spacetime in the dust and Klein-Gordon scalar field models and derive the relevant equations of motion for the background dynamics. In section \ref{sec:numeric},  numerical solutions for some representative initial conditions are discussed when the late-time inflation is driven by a single scalar field with a Starobinsky potential. The emphasis will be placed on the different effects of the reference clocks on the background dynamics due to their distinct evolutionary properties. In section \ref{sec:conclusion}, we  summarize and discuss the main results of the paper.

\section{Review of relational formalism with dust and scalar reference fields}
\label{sec:review}
\renewcommand{\theequation}{2.\arabic{equation}}\setcounter{equation}{0}
In the relational formalism, the Dirac observables are constructed by means of reference fields, whose values at each coordinate point are encoded in a gauge fixing condition.  The observable evaluated at physical coordinates of a given field  returns the value of the field when the  reference fields take those particular values fixed by the gauge conditions. Although one can choose the reference fields from the geometric degrees of freedom \cite{gh2018,ghs2018,gsw2019,tomek-time}, it is often more convenient to consider dust or scalar fields as the reference fields \cite{gtt2010,t2006,ghtw2010I} since the resulting physical  Hamiltonian that generates the dynamics on the reduced phase space is time independent for these models.  Furthermore, these dust and scalar field models have the property that the Poisson algebra of the observables satisfies the standard canonical commutation relation, which is not given in the case of geometric clocks.
In this section, we will review the relational formalism for the Brown-Kucha\v r  and Gaussian dust  reference fields \cite{ghtw2010I,gt2015} as well as a four scalar reference fields \cite{gv2016,gv2017}. As the content in this section has been extensively studied in the vast literature \cite{bergman1961, bergman2, rovelli, komar1958, kuchar1991, dittrich, gh2018, ghs2018, thiemann2006, gt2015, ghtw2010I, ghtw2010II,laura, Vy1994, g2008, rovelli2, dittrich2}, we will only go through the basic ideas and quote the main results and consider their symmetry reduction to a spatially flat FLRW universe as our starting point for the loop quantization. 

\subsection{The relational formalism with  Brown-Kucha\v r, Gaussian dust and scalar reference fields}

The three models considered in this section have the property that they lead to addition of  matter fields to GR coupled to a generic scalar field $\varphi$. In the case of the Brown-Kucha\v{r} and the Gaussian dust model these are 8 additional fields whereas in the case of four scalar field model these are 7 additional fields. Both of the systems are second class systems and the partial reduction of the second class constraints in the individual models leads to a first class system with four additional reference fields. In addition to gravity we consider a generic scalar field $\varphi$ and in the case of the dust model 8 fields denoted by  $T^\mu=(T,S^j,\rho,W_j)$ where $j=1,2,3$ on a four-dimensional hyperbolic spacetime $(\mathcal M, g)$. In the case of the scalar field model we denote the additional scalar fields by $(\chi
^0,\chi^j,M_{jj})$ where again $j=1,2,3$. 
The total action of the coupled system is given by 
\bq
\lb{action}
S=S_\mathrm{geo}+S_\mathrm{scalar}+S_\mathrm{ref}
\eq
where the first two terms of the action are given by
\bqn
S_\mathrm{geo}&=&\frac{1}{2\kappa}\int_{\mathcal M} \der^4x\sqrt{-g} R^{(4)},\\
S_\mathrm{scalar}&=&\frac{1}{\lambda_\varphi}\int_{\mathcal M} \der^4x \sqrt{-g} \left(-\frac{1}{2}g^{\mu \nu} \partial_\mu\varphi \partial_\nu\varphi-U(\varphi)\right),
\eqn
with $\kappa =8\pi G$, $R^{(4)}$ denotes the four-dimensional Ricci scalar, $\lambda_\varphi$ is a coupling constant allowing for a dimensionless $\varphi$ and $U(\varphi)$ is the scalar potential. The last term $S_\mathrm{ref}$ in the action (\ref{action}) depends on the reference field model under consideration and differs for the three models \cite{bk1995,gt2015,gv2016}:
\begin{eqnarray}
\lb{Actref}
S^\mathrm{BK}_\mathrm{dust}&=& -\frac{1}{2}\int_\mathcal M \der^4x\sqrt{-g}\rho[g^{\mu\nu}\widetilde U_\mu\widetilde U_\nu+1], \\
S^\mathrm{G}_\mathrm{dust}&=&-\int \der^3x \sqrt{-g}\left(\frac{\rho}{2}[g^{\mu\nu}T_{,\mu}T_{,\nu}+1]+g^{\mu\nu}T_{,\mu}W_jS^j_\nu\right), \\
S^{\rm ref}_\mathrm{scalar}&=&-\frac{1}{2}\int_\mathcal M \der^4x\sqrt{-g}g^{\mu\nu}\chi^0_{,\mu}\chi^0_{,\nu}
-\frac{1}{2}\int_\mathcal M \der^4x\sqrt{-g}g^{\mu\nu}M_{jj}\chi^j_{,\mu}\chi^j_{,\nu}, 
\end{eqnarray}
with the unit time-like dust velocity field defined by $\widetilde U=-dT+W_j dS^j$, where $j=1,2,3$. The main motivation for introducing only seven but not eight additional scalar fields in the last model comes from the aim to formulate a model close to the model in \cite{dgkl2010} where only one Klein-Gordon scalar field was taken as the temporal reference field but none for the spatial diffeomorphisms. In this way we choose the same temporal reference field as in \cite{dgkl2010}. As shown in \cite{gv2016} considering a model with only four Klein-Gordon scalar fields as reference fields yields a physical Hamiltonian that cannot be quantized in the framework of LQG. 

Let us briefly summarize the main properties of the different models. 
In the Brown-Kucha\v r dust model, as well as the Gaussian dust model, the dust field action involves eight scalar degrees of freedom, collectively denoted by $T^\mu=(T,S^j)$ and $(\rho, W_j)$. For both models adding the dust action to gravity plus standard matter yields a second class system. If we introduce the corresponding Dirac bracket the second class constraints can be solved strongly. In the partially reduced system the four dust fields $T^\mu=(T,S^j)$ are dynamical, whereas the scalar fields $(\rho, W_j)$ can be expressed in terms of the remaining variables in the partially reduced phase space. For the variables on the partially reduced phase space the Dirac bracket coincides with the Poisson bracket. 
The detailed analysis of the Hamiltonian formulation of the dust models \cite{ghtw2010I,gt2015} reveals that, although in the extended phase space, there are 38  local degrees of freedom which include lapse $N$, shift $N^a$, 3-metric $q_{ab}$, the scalar field $\varphi$, the dust fields $T^\mu=(T,S^j)$,  $(\rho, W_j)$ and their respective conjugate momenta, there are also 8 first class constraints and 8 second class constraints which result in a total of 14 physical degrees of freedom in the finally reduced phase space also called physical phase space. In the partially reduced phase space in which only the second class constraints have been implemented the elementary variables  are  the variables $N$,  $N^a$, $q_{ab}$, $\varphi$ and $T^\mu$ and their conjugate momenta and 8 first class constraints. 
These first class constraints consist of four primary constraints and four secondary constraints. The primary constraints are the conjugate momenta of the lapse  function and shift vector,  i.e. $\pi_\mu=(\pi, \pi_a)$. The secondary constraints are the total Hamiltonian and spatially diffeomorphism constraint which take different forms in two dust models that will be discussed separately in the following. Going from the partially reduced phase space to the physical phase space can be achieved by applying the observable map to all elementary variables on the partially reduced extended phase space. The independent physical degrees  of freedom are encoded into the Dirac observables of the variables $(q_{ab},p
^{ab})$ and $(\varphi,\pi_\varphi)$, which are exactly the 14 degrees of freedom mentioned above, whereas the observables associated with $N,N^a$ and their momenta are either phase space independent or can be expressed in terms of these 14 physical degrees of freedom. In the scalar field reference model we start with seven additional scalar fields $(\chi^0,\chi^j,M_{jj})$ next to gravity and a scalar field with potential, so in total with 36 degrees of freedom in phase space. 
Again the resulting system is second class and the 6 second class constraints can be used to eliminate the $M_{jj}$ and their conjugate momenta. In this partially reduced phase space we have 30 degrees of freedom consisting of $N$,  $N^a$, $q_{ab}$, $\varphi$, $\chi^0,\chi^j$ and their conjugate momenta together with 8 first class constraints. The latter include the primary constraints, that is the vanishing of the momenta for lapse and shift, as well as the secondary ones, being the Hamiltonian and spatial diffeomorphism constraints which here involve the corresponding contributions from the scalar reference fields. Now we can proceed as in the dust models and use $\chi^0,\chi^j$ as reference fields and construct Dirac observables. The reduced phase space in this model consists of the Dirac observables associated with 
$(q_{ab},p^{ab})$ and $(\varphi,\pi_\varphi)$, which in this model can also be identified with  the physical degrees of freedom.
In the following, we briefly present the formulas and properties of the two dust models and the four scalar field model that will be relevant for our work in this article: If we perform a partial reduction with respect to the second class constraints, which is in detail discussed in \cite{ghtw2010I,gt2015,gv2016} in all three models,  the Hamiltonian and spatial diffeomorphism constraints can be expressed  as 
\begin{eqnarray}
 \label{eq:DepConstraint}
  c^\mathrm{tot}&=&P^{\rm ref}+h^{\rm ref}, \quad 
 c^{\rm tot}_j=P^{\rm ref}_j+h_j^{\rm ref},
\end{eqnarray}
where the explicit form of $h^{\rm ref}$ and $h_j^{\rm ref}$ depends on the chosen reference model. In the case of the dust models $P^{\rm ref}=P_T$ and $P^{\rm ref}_j=P_j$, whereas for the four scalar field model we have $P^{\rm ref}=\pi_0$ and $P^{\rm ref}_j=\pi_j$, where $P_T$ and $P_j$ are conjugate momenta of $T$ and $S^j$ and $\pi_0$ and $\pi_j$ are the momenta conjugate to $\chi^0$ and $\chi^j$ respectively.
The contributions to the Hamiltonian constraint for the three models are given by
\begin{eqnarray}
 \lb{2a4}
  h^{\rm BK}&=:&-\mathrm{sgn}(P_T)\sqrt{c^2-q^{ab}c_ac_b},\quad h^{\rm G}=c\sqrt{1+q^{ab}T_{,a}T_{,b}}-q^{ab}T_{,a}c_b, \\
  h^{\rm scalar}&=&-\frac{B}{2}-\sqrt{\left(\frac{B}{2}\right)^2-A},
\end{eqnarray}
where 
\begin{eqnarray}
B&:=&-2\sqrt{q}\sum\limits_{j=1}^3\chi^0_{,a}\varphi^a_j\sqrt{q^{cd}\chi^j_{,c}\chi^j_{,d}}, \quad 
A:=qq^{ab}\chi^0_{,a}\chi^0_{,b}-2\sqrt{q}\sum\limits_{j=1}^3\varphi^a_jc_a\sqrt{q^{cd}\chi^j_{,b}\chi^j_{,d}}+2\sqrt{q}c ~.
\end{eqnarray} 
In the above formulas,  $c$ and $c_a$ are given by 
\bqn
\lb{2a2}
c&=&\frac{1}{\sqrt{q}}\left(p_{ab}p^{ab}-\frac{1}{2}(p^{ab}q_{ab})^2\right)-\sqrt{q}R^{(3)}+\frac{\kappa\lambda_\varphi}{\sqrt{q}}\pi^2_\varphi+\frac{2\kappa\sqrt{q}}{\lambda_\varphi}\left(\frac{1}{2}q^{ab}\partial_a \varphi \partial_b \varphi+U\right),\\
\lb{2a6}
c_a&=&-2q_{ab}D_cp^{bc}+2\kappa \pi_\varphi \partial_a \varphi,
\eqn
where $p^{ab}$ and $\pi_\varphi$ are the conjugate momenta of $q_{ab}$ and $\varphi$, respectively, and $q$ is the determinant of the 3-metric. 
The different contributions to the spatial diffeormphism constraints for the individual models read
\bqn
 \lb{2a5}
h^{\rm BK}_j&:=&+S^a_j\left(-h^{\rm BK} T_{,a}+c_a\right),\quad h^{\rm G}_j:=+S^a_j\left(-h^{\rm G} T_{,a}+c_a\right),\nonumber \\
h^{\rm scalar}_j&:=&\chi^a_j\left(-h^{\rm scalar} \chi^0_{,a}+c_a\right), 
 \eqn
where $S^a_j$ is the inverse of $S^j_{,a}$ with $S_{,a}^jS^a_k=\delta^j_k$ and $S^j_{,a}S^b_j=\delta_a^b$ and similarly for $\chi^a_{j}$. In the following we will also use  $c^\mathrm{tot}_\mu=(c^\mathrm{tot},c^\mathrm{tot}_j)$ which collectively denoes the Hamiltonian and diffeomorphism constraint. Next let us introduce the notation to collectively denote the secondary and primary constraints $c_I = (c^\mathrm{tot}_\mu,\pi_\mu)$ and reference fields and their temporal derivatives $T^I=(T^\mu,\dot T^\mu)$ for the dust model and $T^I=(\chi^\mu,\dot \chi^\mu)$  with the multi-index $I=1,\dots 8$. In order that the reference fields can be used to construct Dirac observables, an important condition they need to satisfy is  that $T^I$ and $c_J$ form a canonical conjugate pair, that is,
\bq
\lb{app3}
\{T^I(t,\vec x), c_J(t,\vec y)\}\approx 2\kappa \delta^I_J\delta(\vec x-\vec y). 
\eq
which is satisfied for both dust models as well as the scalar field model. The corresponding gauge fixing conditions for the reference fields are given by
${\mathcal G}^I=({\mathcal G}^\mu,\dot{\mathcal G}^\tau)$ with ${\mathcal G}^\mu=\tau^\mu-T^\mu$ for the dust models and ${\mathcal G}^\mu=\tau^\mu-\chi^\mu$ in the case of the scalar field model, where $\tau^\mu=(\tau, \vec \sigma)$ are  functions on ${\cal M}$ but not dynamical variables on the phase space. With this, one can construct an observable map for an arbitrary function $f$ on the (partially reduced with respect to the second class constraints) extended phase space,  which maps $f$ to its gauge-invariant extension $O_{f,T^\mu}$ and $O_{f,\chi^\mu}$ respectively, that is $f\mapsto {\mathcal O}_{f,T^\mu}(\tau^\mu)$ and performs a reduction with respect to the remaining first class constraints. The explicit form of this map is given by 
\begin{equation}
\label{obsmap}
\begin{split}
 {\mathcal O}_{f,T^\mu}(\tau^\mu)
 &= f+\sum\limits_{n=1}^\infty\frac{1}{n!2^n\kappa^n}\prod\limits_{k=1}^n\int\limits_\sigma d^3x_k{\mathcal G}^J(x_k)\{f(x),{c}_J(x_k)\}_{(n)},
\end{split}    
\end{equation}
where $\{f,g\}_{(n)}$ denotes the iterated Poisson bracket with $\{f,g\}_{(0)}=f$ and $\{f,g\}_{(n)}=\{\{f,g\}_{(n-1)},g\}$. In the case of the scalar field model $O_{f,\chi^\mu}$ can be constructed similarly by replacing the reference fields $T^\mu$ by $\chi^\mu$.
As discussed in detail in \cite{ghtw2010I,gt2015} for the dust models the construction of the observables is performed in two steps. First a reduction with respect to the spatially diffeomorphism constraints is obtained by means of pulling back all variables in the phase space except $(S^j,P_j)$ to the dust manifold yielding  spatially diffeomorphism invariant quantities. As shown in \cite{gv2016} the same strategy can be followed in the scalar field model where all phase space variables but $(\chi^j,\pi_j)$ are pulled back to the scalar field manifold. In the second step in all three models the observable map is applied to these variables to further obtain the reduction with respect to the Hamiltonian constraint and the primary constraints, where the latter has not been discussed in \cite{ghtw2010I,gt2015}, but the explicit form of lapse and shift has been obtained via an alternative route.  The Dirac observable $\mathcal O_{f, T}(\tau)$ and $\mathcal O_{f, \chi^0}(\tau)$ respectively is the image of $f$ under the  observable map. For each $\tau^\mu$,  it returns the value of $f$ in the gauge in which the reference fields $T^\mu$ and $\chi^\mu$ respectively take the values $\tau^\mu$. It is also straightforward to show that $\mathcal O_{f, T}(\tau)$ and $\mathcal O_{f, \chi^0}(\tau)$ weakly Poisson commutes with all the constraints $c_I$ by using the property in (\ref{app3}). The elementary canonical variables in the reduced phase space are made up of the Dirac observables  of  the 3-metric, the scalar field and their respective momenta. More specifically, to keep our notation compact,  we use the notation for these elementary observables to simply denote them by the corresponding capital letters as  
\bq
\lb{app0}
Q_{ij}\coloneqq \mathcal O_{q_{ij}, T (\chi^0)}[\tau, \vec \sigma], \quad P^{ij}\coloneqq \mathcal O_{p^{ij}, T (\chi^0)}[\tau, \vec \sigma], \quad \Phi \coloneqq \mathcal O_{\varphi, T (\chi^0)}[\tau, \vec \sigma], \quad \Pi_\Phi \coloneqq \mathcal O_{\pi_\varphi, T (\chi^0)}[\tau, \vec \sigma]. 
\eq
where the the bracket $(\chi^0)$ indicates that we will use the same capital letter for observables constructed in the dust and scalar field model whenever it is clear from the context which reference fields have been chosen. For all three models these observables 
satisfy the following Poisson brackets 
\bq
\lb{poisson1}
\Big\{Q_{ij}[\tau,\vec \sigma],P^{kl}[\tau,\vec \sigma']\Big\}=2\kappa \delta^k_{(i}\delta^l_{j)}\delta^3(\vec \sigma-\vec \sigma'),\quad \quad
\Big\{\Phi[\tau,\vec \sigma], \Pi_\Phi[\tau,\vec \sigma')\Big\}=\delta^3(\vec \sigma-\vec \sigma').
\eq 
The application of the observable map
to lapse and shift degrees of freedom leads for the Brown-Kucha\v{r} and the Gaussian dust model to the following observables:
\begin{eqnarray}
\lb{2a7}
\mathcal O^{\rm BK}_{N, T}(\tau, \vec \sigma)&=&\frac{C}{H}=\sqrt{1+\frac{Q^{ij}C_iC_j}{H^2}},\quad \mathcal O^{\rm G}_{N, T}(\tau, \vec \sigma)=\frac{C}{H}=\sqrt{1+\frac{Q^{ij}C_iC_j}{H^2}},\nonumber \\
\mathcal O^{\rm BK/G}_{{N}^j, T}(\tau, \vec \sigma)&=&0, \quad \mathcal O^{\rm BK/G}_{\pi, T}(\tau, \vec \sigma)=\pi,\quad 
\mathcal O^{\rm BK/G}_{\pi_j, T}(\tau, \vec \sigma)=\pi_j,
\end{eqnarray}
which shows that in the Brown-Kucha\v{r} as well as the Gaussian model, the lapse and shift degrees of freedom can be completely expressed in terms of the physical variables and the primary constraints and expect for the observables for the lapse $N$ the observables agree in both models\footnote{In \cite{ghtw2010I,ghtw2010II} a shift vector in terms of physical degrees of freedom is defined as $N^j=-\frac{Q^{jk}C_k}{H}$ whereas in the above we present the Dirac observable associated to the shift vector which is just zero. Note that there is no contradiction between  two results because the $N
^j$ from \cite{ghtw2010I,ghtw2010II} is not the Dirac observable of the shift vector but has been defined as follows. Consider $\{F,H(\sigma)\}=\frac{C}{H}\{F,C\}-\frac{Q^{jk}C_k}{H}\{F,C_j\}$. Then $N^j$ was defined as being the coefficient in front of the Poisson bracket of $\{F,C_j\}$.}.
On the other hand for the scalar field model these observables take the following form
 \bq
\lb{eq:ScalNNj}
\mathcal O^{\rm scalar}_{N, \chi^0}(\tau, \vec \sigma)=-\frac{\sqrt{Q}}{H^{\rm scalar}},\quad \mathcal O^{\rm scalar}_{{N}^j, \varphi}(\tau, \vec \sigma)=\frac{\sqrt{Q}\sqrt{Q^{jj}}}{H^{\rm scalar}}, \quad \mathcal O^{\rm scalar}_{\pi, \chi^0}(\tau, \vec \sigma)=\pi,\quad 
\mathcal O^{\rm scalar}_{\pi_j, \chi^0}(\tau, \vec \sigma)=\pi_j,
\eq
In the reduced phase the evolution of a generic phase space function $F(Q_{ij}, P^{ij}, \Phi, \Pi_\Phi)$ is governed by Hamilton's equations that involve the so called physical Hamitlonian ${\bf H}_\mathrm{phys}$
\bqn
\lb{ham}
\dot F&=&\{F, {\bf H}_\mathrm{phys}\}
=\int_\mathcal S\der^3\sigma\{F, H(\sigma)\}.
\eqn
The corresponding Hamiltonian density $H(\sigma)$ of the physical Hamiltonian is the observable associated the function $h^{\rm ref}$ in (\ref{eq:DepConstraint}) and consequently differs for each of the three models. The physical Hamiltonian associated with for the Brown-Kucha\v{r} model using ${h}^{\rm BK}$ takes the form
\bq
\lb{bkham}
{\bf H}^{\rm BK}_\mathrm{phys}=\frac{1}{2\kappa}\int\limits_{\cal S} \der^3\sigma\,
O_{{h}^{\rm BK},T}
=\frac{1}{2\kappa}\int\limits_{\cal S}\, \der^3\sigma\,\sqrt{C^2-Q^{ij}C_iC_j}=:\frac{1}{2\kappa}\int\limits_{\cal S}\, \der^3\sigma\, H^{\rm BK}(\sigma) .
\eq
Here $H^{\rm BK}:=\sqrt{C^2-Q^{ij}C_iC_j}$ denotes the physical Hamiltonian density in the Brown-Kucha\v{r} model. $C$ and $C_i$ are the images of (\ref{2a2})-(\ref{2a6}) under the observable map which are explicitly given by  
\bq
\lb{2a3}
C=\frac{1}{\sqrt{Q}}G_{ijmn}P^{ij}P^{mn}-\sqrt{Q}R^{(3)}+\frac{\kappa\lambda_\varphi\Pi^2_\Phi}{\sqrt{Q}}+\frac{2\kappa\sqrt{Q}}{\lambda_\varphi}\left(\frac{1}{2}Q^{ij}\partial_i \Phi \partial_j \Phi+U\right),
\eq
and 
\bq
\lb{2a8}
C_i=-2 D_k P^k_i+2\kappa\Pi_\Phi \partial_i \Phi.
\eq
It should be noted that in order to obtain the Hamiltonian (\ref{bkham}), we choose $\mathrm{sgn}(P_T)$ to be negative so that (\ref{bkham}) is bounded from below.
In the case of the Gaussian model we obtain for the physical Hamiltonian
\bq
\lb{gaussham}
{\bf H}^{\rm G}_\mathrm{phys}=\frac{1}{2\kappa}\int\limits_{\cal S}\der^3\sigma\,
{\cal O}_{{h}^{\rm G},T}=\frac{1}{2\kappa}\int\limits_{\cal S} \der^3\sigma\, C,\quad C:={\cal O}_{c,T}=c(Q^{ij},P_{ij},\Phi,\Pi_\Phi),
\eq
The four scalar field model has a physical Hamiltonian of the form 
\bqn
\lb{eq:Scalham}
{\bf H}^{\rm scalar}_\mathrm{phys}&=&:\frac{1}{2\kappa}\int\limits_{\cal S} \der^3\sigma\,
{\cal O}_{{h}^{\rm scalar},\chi^0}
=:\frac{1}{2\kappa}\int\limits_{\cal S}\, \der^3\sigma\,\sqrt{-2\sqrt{Q}C+2\sqrt{Q}\sum\limits_{j=1}^3\sqrt{Q^{jj}C_jC_j}}
\eqn
 where $C$ and $C_j$ take the same form as in (\ref{2a3}) and (\ref{2a8}) and we define the Hamiltonian density  $H^{\rm scalar}(\sigma):=\sqrt{-2\sqrt{Q}C+2\sqrt{Q}\sum\limits_{j=1}^3\sqrt{Q^{jj}C_jC_j}}$. A difference to the  dust models considered before is that here summation over $j$ involved in the physical Hamiltonian density induces a kind of directional dependence at the level of the scalar field manifold for the reason that the $C_j$'s and the inverse metric $Q^{jk}$ are not contracted in a covariant fashion. However, since this is at the level of the observables and refers to the scalar field manifold, it causes no issue here. Moreover, in the context of this work where we consider the FLRW symmetry reduced case this term in the physical Hamiltonian will not contribute but vanishes in this sector. The reason for the form of this term can be understood when looking at the explicit form of the Dirac observables for lapse and shift and their conjugate momenta in this model in (\ref{eq:ScalNNj}).
 %%%%%%%%
 %%%%%%%%
 \subsection{The reduced phase space in a spatially flat FLRW universe}
 \label{sec:RedPSFLRW}
In this section we want to take the reduced phase spaces obtained in three models as a starting point and further symmetry reduce it to the case of a spatially flat FLRW universe. For the reason that we want to apply a reduced phase space quantization in the context of loop quantum cosmology later on we will consider the reduced phase spaces formulated in terms of Ashtekar-Barbero variables. If we perform the extension from the ADM phase space to the one where the Ashtekar-Barbero variables $(A_a^i,E^a_i)$ are the elementary ones we obtain an additional SU(2) Gauss constraint, which for the full theory will be handled using Dirac quantization. If we restrict to the FLRW symmetry reduced sector the Gauss constraint plays anyway no role for our work presented here. Following the discussion of the former section we use in each of the presented models the reference fields and the observable map to obtain the Dirac observables of the (up to the Gauss constraint) reduced phase space that are for the two dust models  denoted by
\bq
\lb{eq:obsAE}
 \mathcal O_{A^i_a, T}[\tau, \vec \sigma],\quad \mathcal O_{E^a_i,T}[\tau, \vec \sigma], \quad  O_{\varphi, T}[\tau, \vec \sigma], \quad \mathcal O_{\pi_\varphi, T}[\tau, \vec \sigma], 
\eq
and in the case of the four scalar field model are given by 
\bq
\lb{eq:obsAEScal}
\mathcal O_{A^i_a, \chi^0}[\tau, \vec \sigma],\quad  \mathcal O_{E^a_i,\chi^0 }[\tau, \vec \sigma], \quad  O_{\varphi, \chi^0}[\tau, \vec \sigma], \quad \mathcal O_{\pi_\varphi, \chi^0}[\tau, \vec \sigma] .
\eq
The corresponding physical Hamiltonians ${\bf H}^{\rm G}_{\rm phys},{\bf H}^{\rm BK}_{\rm phys}$ and ${\bf H}^{\rm scalar}_{\rm phys}$ are then understood as function of these Dirac observables.

If we specialize to the symmetry reduced FLRW sector we consider a spatially flat isotropic and homogeneous FLRW spacetime described by the metric
\bq
\lb{3a1}
\der s^2=-\mathcal{O}_{N^2} \der \tau^2+ Q_{jk} \der \sigma^j \der \sigma^k,\quad {\rm and}\quad 
\der s^2=-\mathcal{O}_{N^2} \der \tau_\chi^2+ Q_{jk} \der \sigma^j \der \sigma^k,
\eq
respectively, where $\tau$ is the physical time of the dust models that can be interpreted as proper time and $\tau_\chi$ denotes the physical time in the scalar field model, $\mathcal{O}_N$ is the lapse function and $Q_{jk}$ is the physical metric related to comoving coordinates as 
\be
Q_{jk}  \der \sigma^j \der \sigma^k= \mathcal{O}_{a^2} \mathring Q_{jk} \der \sigma^j \der \sigma^k=
\mathcal{O}_{a^2}((\der\sigma^1)^2+(\der\sigma^2)^2+(\der\sigma^3)^2).
\ee
Here $\mathring Q_{jk}$ is the fiducial metric over the spatial manifold which for $k=0$ model can be 
$\mathbb{R}^3$ or $\mathbb{T}^3$. In the latter case, the scale factor $a$ relates the coordinate volume $\mathcal{O}_{V_o}$ of the 3-torus to the physical volume as $\mathcal{O}_{V}= \mathcal{O}_{a^3 V_0}$. If the spatial manifold is chosen with spatial topology  $\mathbb{R}^3$ then one needs to choose a fiducial cell ($\mathcal V$), acting as an infra-red regulator, to define symplectic structures. In this case, one is free to choose another cell which has a rescaled coordinate volume. A key requirement for consistency of physics in LQC is that physical predictions of observables which are classically invariant under rescalings of the fiducial cell must be invariant under such rescalings after quantization. It turns out that this requirement leads to so-called improved dynamics or the $\bar \mu$-scheme as the only consistent quantization in isotropic LQC \cite{cs08}. In our analysis we will focus on the $\bar \mu$-scheme introduced in \cite{aps2006c} in the next section.  

Due to the homogeneity and isotropicity of the spatially flat FLRW universe, the reduced phase space of the geometric sector  is a two-dimensional space spanned by  the canonical pair $(\mathcal{O}_c,  \mathcal{O}_p)$ which are related with the Dirac observables of the Ashtekar-Barbero SU(2) connection $\mathcal{O}_{A^i_a}$ and the densitized triad $\mathcal{O}_{E^a_i}$ via
\bq
\lb{3a2}
\mathcal O_{A^i_a}=\mathcal O_c \mathcal O_{V^{-1/3}_0}  \mathcal O_{\mathring \omega^i_a}, \quad \quad  \mathcal O_{E^a_i}= \mathcal O_p \mathcal O_{V^{-2/3}_0} \sqrt{{\mathcal O}_{\mathring q}} \mathcal O_{\mathring e^a_i},
\eq
with $ \mathcal O_{\mathring e^a_i}$ and $\mathcal  O_{\mathring \omega^i_a}$ representing the fiducial triads and co-triads compatible with the fiducial metric $\mathring Q_{jk}$, respectively. The fiducial volume of $\mathcal V$ with respect to the fiducial metric $\mathring Q_{jk}$ will be set to unity in the following. The canonically conjugate pair $(\mathcal{O}_{c}, \mathcal{O}_{p})$ satisfies in all three models 
\bq
\lb{3b2}
\{\mathcal O_c,\mathcal O_p\}=8\pi G \gamma/3 .  
\eq
 Here $\gamma$ is the Barbero-Immirzi parameter which is determined  from black hole thermodynamics in LQG. Its exact value  depends on the prescription chosen to count the degrees of freedom on an isolated horizon. In LQC literature, this value is  conventionally set to $\gamma\approx 0.2375$ which will be considered in this manuscript. Note that here and in the following the subscript $c$ in ${\cal O}_c$ denotes the symmetry reduced connection and does not correspond to the $c$ in (\ref{2a2}) that denotes all but the reference contribution to the Hamiltonian constraint. Since we work in the reduced phase space, in the following the symmetry reduced connection is always depicted in its Dirac observable form and to avoid confusion the Dirac observable associated with $c$ in (\ref{2a2}) is denoted by $C$. For the above equation,  we used that the Dirac bracket of the gauge variant connection and triad variables coincides with their Poisson bracket and the same is true for the matter sector, that is $\{\mathcal O_\varphi, \mathcal O_{\pi_\varphi}\}=1,$ where  we will consider an inflationary field $\varphi$ in Starobinsky inflationary potential (\ref{star}) and $\mathcal O_\varphi , O_{\pi_\varphi}$ denotes the Dirac observables associated with $\varphi,\pi_\varphi$. As before we again use the property that for the variables $\varphi,\pi_\varphi$ we have  $\{\varphi,\pi_\varphi\}^*=\{\varphi,\pi_\varphi\}$, where the star on the right upper corner of the Poisson bracket stands for the Dirac bracket of two variables. 
Now it remains to discuss the symmetry reduced form of the physical Hamiltonians for each of the models. 
In the case of the spatially flat FLRW spacetime,  the observable $C_j$  in (\ref{2a8}) vanishes  and  as a result the physical Hamiltonians in the Brown-Kucha\v r  and Gaussian dust models, which are  given respectively in (\ref{bkham}) and (\ref{gaussham}), coincide. This fits well with the fact that in the case of spatially flat FLRW spacetime the Dirac observable of the lapse function $\mathcal{O}_{N,T}$ in (\ref{2a7}) of the Brown-Kucha\v{r} model becomes unity and thus agrees with the corresponding Dirac observable of the Gaussian dust model. Since in the Brown-Kucha\v r dust model the physical Hamiltonian involves a square root, we briefly discuss this in a bit more in detail for this model. Considering the Hamiltonian constraint in (\ref{2a4}),  in the symmetry reduced case we obtain %for the reduced phase space 
\bq
P_T = {\rm sgn}(P_T)\sqrt{C^2} = {\rm sgn}(P_T)|C|,
\eq
and the physical Hamiltonian will be given by ${\bf H}^\mathrm{FLRW, dust}_\mathrm{phys}=-{\rm{ sgn}}(P_T) |C|$. Now from the original form of the constraints in the Brown-Kucha\v r dust model \cite{ghtw2010I} we have $c^{\rm tot}=c+P\sqrt{1+q^{ab}U_aU_b}$, where $c$ is shown in (\ref{2a2}) and this implies at the level of observables that on the constraint surface ${\rm sgn}(c)=-{\rm sgn}(P_T)$ from which we obtain ${\rm sgn}(C)=-{\rm sgn}(P_T)$. Considering this let us discuss the two possible cases: if the dust energy is negative, that is $P_T<0$, then we have ${\bf H}^\mathrm{FLRW, dust}_\mathrm{phys}=|C|=C$ where we used in the last step that for $P_T<0$ we have $C>0$. If the dust energy density is chosen to be positive, that is $P_T>0$ we have $C<0$ and thus ${\bf H}^\mathrm{FLRW, dust}_\mathrm{phys}=-|C|=C$ in this part of the phase space. Thus we fix the sign of $C$ at the classical level and then quantize the corresponding sectors of the reduced phase space. As discussed in \cite{gt2015} this can become problematic since it requires to have sufficient control on the spectrum of the physical Hamiltonian which for full LQG is perhaps a  too strong requirement. However, in the mini-superspace models considered here we assume that this assumption is justified.

 In order to make the integral in the physical Hamiltonian finite, one can first choose a fiducial cell with the volume $\mathcal O_{\mathcal V_0}$ in the symmetry reduced dust manifold, then compute all the integrals within the fiducial cell. As mentioned earlier, we choose the volume of this cell to be unity. Finally, the physical Hamiltonian of the dust models in the classical theory can be written in terms of the Dirac observables for the symmetry reduced variables as  
\bq
\lb{phyham}
{\bf H}^\mathrm{FLRW, dust}_\mathrm{phys}=\frac{-3\mathcal O^2_c\sqrt{|\mathcal O_p|}}{\kappa \gamma^2}+\frac{\lambda_\varphi \mathcal O^2_{\pi_\varphi}}{2|\mathcal O_p|^{3/2}}+\frac{|\mathcal O_p|^{3/2}U(\mathcal O_\varphi)}{\lambda_\varphi}.
\eq
It should also be noted that ${\bf H}^\mathrm{FLRW, dust}_\mathrm{phys}$ is the physical Hamiltonian in the reduced phase space so it  does not vanish but is a constant of motion which is determined by the initial conditions of the system. 

In the case of the scalar field reference model the symmetry reduction of the physical Hamiltonian leads to the following form
\bq
\lb{phyhamScal}
{\bf H}^\mathrm{FLRW, scalar}_\mathrm{phys}=\sqrt{-2|\mathcal O_p|^{3/2}\left(\frac{-3\mathcal O^2_c\sqrt{|\mathcal O_p|}}{\kappa \gamma^2}+\frac{\lambda_\varphi \mathcal O^2_{\pi_\varphi}}{2|\mathcal O_p|^{3/2}}+\frac{|\mathcal O_p|^{3/2}U(\mathcal O_\varphi)}{\lambda_\varphi}\right)}.
\eq
This setup will be taken as the starting point of the reduced phase space quantization of the symmetry reduced flat FLRW sector for the three models in the next section.

\section{Loop quantization of the spatially flat universe in the reltional formalism}
\label{sec:LQC}
\renewcommand{\theequation}{3.\arabic{equation}}\setcounter{equation}{0}
In this section, we apply loop quantization to the reduced phase space of the dust models and Klein-Gordon scalar field model for a spatially flat FLRW universe. Then, we will proceed with the effective description of quantum spacetime  and find the resulting Hamilton's equations and modified Friedmann equation from an effective Hamiltonian. Then we perform numerical analysis using Hamilton's equations in the next section. 
Let us note that conventional LQC allows different factor orderings, such as \cite{aps2006c,acs2010,mmo}. While we present results following the construction in \cite{aps2006c}, which also guides our notation, our  results can be generalized in a straightforward way to other factor orderings. 
%%%%%%

For all three reference field models we consider the reduced phase space symmetry reduced to a spatially flat FLRW universe discussed in section \ref{sec:RedPSFLRW} as the starting point for quantization. What we are aiming at is to find a representation of the algebra of Dirac observables for each individual model that further allows to implement the physical Hamiltonian of the models as a well defined operator.  In this approach we directly obtain a representation of the physical Hilbert space. This is in contrast to the usual approach in LQC with a scalar field clock. There one deparameterizes the Hamiltonian constraint at the classical level and then performs a Dirac quantization by promoting the classical constraint to a constraint operator on the kinematical Hilbert space. Considering then states that are annihilated by the constraint yields a set of functions on which the physical inner product can be defined (see for instance \cite{aps2006b,aps2006c}), 
and taking the completion with respect to the corresponding norm yields the physical Hilbert space in this route.

An advantage of all these models have is that the algebra of the elementary Dirac observables in the reduced phase space is as simple as the kinematical algebra, that is 
\bq
\{\mathcal{O}_{c},\mathcal{O}_{p}\} =\frac{8\pi G\gamma}{3} ,\quad  \{\mathcal{O}_\varphi,\mathcal{O}_{\pi_\varphi}\} =1,    
\eq
where we as before neglected the label of the clocks here that would be $T$ for the dust models and $\chi^0$ for the scalar model.  Since we want to apply a  loop quantization, instead of the connections $\mathcal{O}_c$ and triads $\mathcal{O}_{p}$, the canonical variables used for quantization are the holonomies of the connection along edges and the fluxes of the triads along 2-surfaces. These can be constructed from our Dirac observables  $\mathcal{O}_c$ as follows:
\bq
\lb{3a4}
{\cal O}_{h^{(\mu)}_k}=\cos\left(\frac{\mu {\cal O}_c}{2}\right)\mathbb{1}+2\sin\left(\frac{\mu {\cal O}_c}{2}\right)\tau_k,
\eq
where $\mathbb{1}$ represents a unit $2\times2$ matrix and $\tau_k=-i\sigma_k/2$ with $\sigma_k$ denoting the Pauli spin matrices. To obtain the above equation we have used that the observable map is a homomorphism with respect to the groups of multiplication and addition, that is for any functions $f(c,p,\varphi,\pi_\varphi),$ in the original kinematical phase space, the observable map has the property
\bq
\lb{3b3}
\mathcal O_{f(c,p,\varphi,\pi_\varphi),T}=f(\mathcal{O}_{c,T},\mathcal{O}_{p,T},\mathcal{O}_{\varphi,T},\mathcal{O}_{\pi_\varphi,T})
\quad{\rm and}\quad 
\mathcal O_{f(c,p,\varphi,\pi_\varphi),\chi^0}=f(\mathcal{O}_{c,\chi^0},\mathcal{O}_{p,\chi^0},\mathcal{O}_{\varphi,\chi^0},\mathcal{O}_{\pi_\varphi,\chi^0}),
\eq
in the two dust models and in the scalar field model respectively. 

The flux of the triads along 2-surfaces in the spatially flat FLRW spacetime turns out to be proportional to\footnote{For an alternate quantization using gauge-covariant fluxes see \cite{ls19}.} $\mathcal O_p$. The elementary variables that are chosen instead of $\mathcal{O}_c,\mathcal{O}_{p}$ in LQC are the pair $N_\mu(\mathcal{O}_c)=e^{i\mu \mathcal O_c/2}$ and $\mathcal{O}_p$, where $\mu\in\mathbb{R}$ that satisfy the algebra
\bq
\{N_\mu(\mathcal{O}_c),\mathcal{O}_p\}= i \frac{4 \pi G \gamma \mu}{3} N_\mu(\mathcal{O}_c). 
\eq
Because the algebra of the elementary variables on our reduced phase space coincides exactly with the one at the kinematical level of LQC we can use the representation usually describing the kinematical Hilbert space of LQC ${\cal H}_{\rm kin}$ here as a representation of the physical Hilbert space ${\cal H}_{\rm phys}$. This is in exact analogy to the model presented in \cite{gt2007,gv2016,gv2017} based on a reduced quantization of full LQG. 
%%%%%%%%%%%%
\subsection{Physical Hilbert space and Hamiltonian operator of the dust models} 
As far as the gravitational sector of the theory is considered thanks to the methods in LQG \cite{lost,flost}, a unique representation for the symmetry reduced model can be identified \cite{unique-lqc-rep1,unique-lqc-rep2}, which can be used for the physical Hilbert space for the reference field models in our work. In the case of the two dust models the physical Hilbert space is $\mathcal H_\mathrm{phys, grav}=L^2(\mathbb{R}_\mathrm{Bohr},\der \mu_\mathrm{Bohr})$ where $\mathbb{R}_\mathrm{Bohr}$ is the Bohr compactification of the real line with Haar measure $\mu_\mathrm{Bohr}$.  The sector of the inflaton is quantized in the usual Schr\"odinger representation with the Hilbert space ${\cal H}_{\rm phys,\varphi}=L^2(\mathbb{R}, \der \mathcal{O}_\varphi)$. The physical Hilbert space of the theory is then just given by the tensor product of the individual Hilbert spaces ${\cal H}_{\rm phys}={\cal H}_{\rm phys,grav}\otimes{\cal H}_{\rm phys,\varphi}$. We denote the elements of ${\cal H}_{\rm phys}$ by $\Psi(\mathcal{O}_c,\mathcal{O}_\varphi):=\psi(\mathcal{O}_c)\otimes\psi(\mathcal{O}_\varphi)$. The inner product of ${\cal H}_{\rm phys}$ is given by 
\begin{equation}
\langle \Psi, \widetilde{\Psi}\rangle_{\rm phys}:=\langle \psi,\widetilde{\psi}\rangle_{\rm grav}\langle \psi,\widetilde{\psi}\rangle_{\varphi},
\end{equation}
with the inner product in the gravitational sector 
\begin{equation}
 \langle \psi,\tilde{\psi}\rangle_{\rm grav}:=\lim\limits_{D\to\infty}\frac{1}{2D}\int\limits_{-D}^D \der\mathcal{O}_c \overline{\psi}(\mathcal{O}_c)\widetilde{\psi}(\mathcal{O}_c),
\end{equation}
and for the matter part we  use the standard Schr\"odinger inner product given by
\begin{equation}
 \langle \psi,\tilde{\psi}\rangle_{\varphi}:=\int\limits_{-\infty}^\infty \der\mathcal{O}_\varphi \overline{\psi}(\mathcal{O}_\varphi)\widetilde{\psi}(\mathcal{O}_\varphi).
\end{equation}
If we choose some orthonormal basis in ${\cal H}_{\rm phys,\varphi}$ that we denote by $\psi_n$ with $n\in\mathbb{N}$ then an orthonormal basis in ${\cal H}_{\rm phys}$ is given by the set $(N_\mu(\mathcal{O}_c)\otimes\psi_n)_{\mu\in\mathbb{R},n\in\mathbb{N}}$ since $\langle N_\mu(\mathcal{O}_c)| N_{\mu'}(\mathcal{O}_c) \rangle_{\rm grav} = \delta_{\mu,\mu'}$ where $N_\mu(\mathcal{O}_c)$ are almost periodic functions of $\mathcal{O}_c$. Normalizable quantum states are given by a tensor product  of a discrete sum of plane waves, that is $\psi({\cal O}_c) = \sum_i a_i e^{i \mu_i {\cal O}_c/2}$ and matter states of the form $\psi(\mathcal{O}_\varphi)=\sum\limits_{n=0}
^\infty \alpha_n \psi_n$ with $\sum\limits_{n=0}^\infty |\alpha_n|^2<\infty$.

Let us briefly discuss how the elementary operators act on the physical Hilbert space ${\cal H}_{\rm phys}$. 
The operators corresponding to the Dirac observables of the  holonomy and fluxes  denoted by $\hat{\mathcal{O}}_{N_\mu}$ and $\widehat{\mathcal{O}}_p$ respectively act on the states $\Psi({\cal O}_c,\mathcal{O}_\varphi)$ by multiplication and differentiation respectively on ${\cal H}_{\rm phys,grav}$, and trivially on ${\cal H}_{\mathrm{phys},\varphi}$:
\begin{eqnarray}
\label{eq:ActionOpgrav}
\widehat {\cal O}_{N_\mu}\Psi({\cal O}_c,\mathcal{O}_\varphi)&:=&(\widehat {\cal O}_{N_\mu}\otimes\mathbb{1}) \Psi({\cal O}_c,\mathcal{O}_\varphi) = e^{{i\mu {\cal O}_c/2}}\psi(\mathcal{O}_c)\otimes \psi(\mathcal{O}_\varphi), \nonumber \\
\widehat {\cal O}_{p} \Psi({\cal O}_c,\mathcal{O}_\varphi)&:=&(\widehat {\cal O}_{p}\otimes\mathbb{1}) \Psi({\cal O}_c,\mathcal{O}_\varphi)=
 - i \frac{8 \pi \gamma \lp^2 }{3}\frac{\der}{\der {\cal O}_c} \psi({\cal O}_c)\otimes\psi(\mathcal{O}_\varphi).
\end{eqnarray}
Likewise the elementary operators corresponding to the Dirac observables $\mathcal{O}_\varphi,\mathcal{O}_{\pi_\varphi}$ which we denote by $\widehat{\mathcal{O}}_\varphi$ and $\widehat{\mathcal{O}}_{\pi_\varphi}$ respectively act trivially on ${\cal H}_{\rm phys,grav}$ and as multiplication and derivative operators respectively on ${\cal H}_{\rm{phys},\varphi}$, with an explicit action of the form:  
\begin{eqnarray}
\label{eq:ActionOpphi}
\widehat {\cal O}_{\varphi}\Psi({\cal O}_c,\mathcal{O}_\varphi)&:=&(\mathbb{1}\otimes\widehat{O}_\varphi) \Psi({\cal O}_c,\mathcal{O}_\varphi) = \psi(\mathcal{O}_c)\otimes \mathcal{O}_\varphi \psi(\mathcal{O}_\varphi), \nonumber \\
\widehat {\cal O}_{\pi_\varphi} \Psi({\cal O}_c,\mathcal{O}_\varphi)&:=&(\mathbb{1}\otimes\widehat{O}_{\pi_\varphi}) \Psi({\cal O}_c,\mathcal{O}_\varphi)=
\psi({\cal O}_c)\otimes \frac{\hbar}{i}\frac{\der }{\der \mathcal{O}_\varphi}\psi(\mathcal{O}_\varphi) ~.
\end{eqnarray}
In the following, as  often done in physics notation, we will suppress the tensor product and  work with the notation on the left hand side of the equations above and the action of the corresponding operators is always understood in the sense defined above.

As in unreduced LQC, we would be working with the triad (or volume) representation in which
 the operator $\widehat {\cal O}_{N_\mu}$ acts as a translation operator
\be
\widehat {\cal O}_{N_{\mu'}} \Psi({\cal O}_p\mathcal{O}_\varphi) = \Psi({\cal O}_{p + \mu'},\mathcal{O}_\varphi) = \Psi({\cal O}_p + \mu',\mathcal{O}_\varphi),
\ee
and $\widehat {\cal O}_p$ acts as a multiplication operator
\be
\widehat {\cal O}_p \Psi({\cal O}_p,\mathcal{O}_\varphi) = \frac{8 \pi \gamma \lp^2}{6} \mu \Psi({\cal O}_p,\mathcal{O}_\varphi) ~.
\ee
Note that since there are no fermions in our analysis, as in LQC, we consider  states $\Psi({\cal O}_p,\mathcal{O}_\varphi)$  such that they are symmetric under parity operation: 
$\widehat \Pi \Psi({\cal O}_p,\mathcal{O}_\varphi) = \Psi(-{\cal O}_p,\mathcal{O}_\varphi)$, i.e. they satisfy $\Psi({\cal O}_p,\mathcal{O}_\varphi) = \Psi(-{\cal O}_p,\mathcal{O}_\varphi)$.

The gravitational part of the Hamiltonian, denoted by ${\bf H}^{\rm grav}_\mathrm{phys}$, is obtained by expressing terms involving the symmetry reduced connection  in terms of holonomies along the edges of a square with comoving length $\ell \mathcal{O}_{V_0^{1/3}}$. In terms of the Dirac observables, the gravitational part in the classical theory can be written as 
\bq
\lb{3a8}
{\bf H}^{\rm grav}_\mathrm{phys} = \lim_{\ell \rightarrow 0} \sin\left(\ell {\cal O}_c \right)\Big[-\frac{\mathrm{sgn}({\cal O}_p)}{32\pi^2 G^2 \gamma^3\ell^3}\sum_k \mathrm{Tr} \tau_k {\cal O}_{h^{(\ell)}_k}\{{\cal O}_{h^{(\ell)}_k}^{-1},{\cal O}_{|p|}^{3/2}\}\Big]\sin\left(\ell {\cal O}_c\right).
\eq
Here in contrast to unreduced LQC \cite{aps2006c},  
${\bf H}^{\rm grav}_\mathrm{phys}$ is expressed in terms of Dirac observables. In particular, ${\bf H}^{\rm grav}_\mathrm{phys}$ is the Dirac observable of the rescaled gravitational contributions to the Hamiltonian constraint denoted by ${C}_\mathrm{grav}/16 \pi G$ in \cite{aps2006c}. We will denote the corresponding operator on the physical Hilbert space ${\cal H}_{\rm phys}$ by $ \widehat {\bf H}^{\rm grav}_\mathrm{phys}$ which is a true Hamiltonian in our approach, whose action on physical states  {\it{does not vanish}} in the physical Hilbert space. In contrast to this in \cite{aps2006c} Dirac quantization was applied for the $k=0$ FLRW model sourced with a massless scalar field used as a relational clock and consequently the quantization of ${C}_\mathrm{grav}/16 \pi G$ was performed which vanishes in the quantum theory.

Let us recall that in LQC, $\ell$ is chosen to coincide with the physical length of the square loop with area given by the minimum area eigenvalue in LQG. In the improved dynamics prescription \cite{aps2006c}, this physical length is measured by $\bar \mu=\sqrt{\Delta/{\cal O}_{|p|}}$ with $\Delta=4\sqrt{3}\pi\gamma l^2_\mathrm{Pl}$. Unlike other possible prescriptions, the improved dynamics is the only known choice which yields an unambiguous scale for bounce density and an infra-red limit which is free from rescalings of the fiducial cell \cite{cs08}. In this prescription, since $\bar \mu$  depends on the triad $\mathcal{O}_p$, the holonomy operator ${\cal O}_{h^{(\bar \mu)}_k}$ when restricted to ${\cal H}_{\rm phys,grav}$ is  the shift operator on eigenstates of the volume operator $|\nu\rangle$ where $\nu= K\mathrm{sgn}(\mu)|\mu|^{3/2}$ with $K=\frac{2\sqrt{2}}{3\sqrt{3\sqrt{3}}}$ and acts trivially on ${\cal H}_{\rm{phys},\varphi}$. For this reason, it is convenient to switch from the triad to the volume representation in ${\cal H}_{\rm phys,grav}$ (see \cite{aps2006c} for more details). In this case 
\begin{equation}
{\cal H}_{\rm phys,grav}=\overline{{\rm span}(|\nu\rangle \, :\, \nu\in\mathbb{R})}  \quad{\rm with}\quad 
\langle \nu\, , \, \nu'\rangle = \delta_{\nu \nu'}.
\end{equation}
When acting on the volume-kets in ${\cal H}_{\rm phys,grav}$, the action of the operator corresponding to the Dirac observables for the volume, defined as $\widehat {\cal O}_V=\widehat{{{\cal O}_{|p|}}^{3/2}}$, and holonomy operators is given by
\bq
\lb{3a10}
\widehat {\cal O}_V |\nu\rangle=\left(\frac{8\pi G\gamma}{6}\right)^{3/2}\frac{|\nu|}{K}|\nu\rangle, \quad \quad \widehat {\cal O}_{N_{\bar \mu}}|\nu\rangle=|\nu+1\rangle.
\eq
Further, the holonomy operator acts trivially on ${\cal H}_{\rm phys,\varphi}$.

The corresponding quantum operator for the gravitational part of the physical Hamiltonian acting on ${\cal H}_{\rm phys}$, following the construction in \cite{aps2006c}, becomes
\bqn
\lb{3b5}
 \widehat {\bf H}^{\rm grav}_\mathrm{phys}\Psi(\mathcal{O}_V,\mathcal{O}_\varphi)
 &:=&\left(\widehat {\bf H}^{\rm grav}_\mathrm{phys}\otimes\mathbb{1}\right)\Psi(\mathcal{O}_V,\mathcal{O}_\varphi) \nonumber \\
&=&
 \left(\sin \left(\bar \mu {\cal O}_c\right) \Big[ \frac{3 i\mathrm{sgn}({\cal O}_V)}{16\pi^2 G^2\gamma^3\bar \mu^3}\left( \sin \left(\frac{\bar \mu {\cal O}_c}{2}\right) \widehat {\cal O}_V \cos \left(\frac{\bar \mu {\cal O}_c}{2}\right)\right.\right.\nb\\
 &&\left.\left.~~~-\cos \left(\frac{\bar \mu {\cal O}_c}{2} \right) \widehat {\cal O}_V \sin \left(\frac{\bar \mu {\cal O}_c}{2}\right) \right) \Big]\sin\left(\bar \mu {\cal O}_c\right)\right)
 \psi(\mathcal{O}_V)\otimes\psi(\mathcal{O}_\varphi).
\eqn
The quantization of the matter part of the Hamiltonian $\widehat {\bf H}^{\varphi}_{\rm phys}$ with a scalar field $\varphi$ in an inflationary potential $U(\mathcal{O}_\varphi)$ yields
\begin{eqnarray}
\widehat {\bf H}^{\varphi}_{\rm phys}\Psi(\mathcal{O}_V,\mathcal{O}_\varphi)&=&
\left(-\frac{\hbar^2}{2} \widehat{{{\mathcal O}^{-3/4}_{|p|}}} \partial^2_{\mathcal{O}_\varphi} \widehat{{{\mathcal O}^{-3/4}_{|p|}}}+\widehat{{\mathcal O}_{|p|}^{3/4}}U({\widehat{\mathcal O}}_\varphi)\widehat{{\mathcal O}_{|p|}^{3/4}}\right)\Psi(\mathcal{O}_V,\mathcal{O}_\varphi),
\end{eqnarray}
where we chose a symmetric operator ordering.

Given these two individual contributions, the physical Hamiltonian in both dust models is just given by
\be
\widehat{\bf H}_{\rm phys}=\widehat{\bf H}_\mathrm{phys}^\mathrm{grav}+\widehat{\bf H}_{\rm phys}^\varphi ~.
\ee
At the classical level in the reduced phase space ${\bf H}_{\rm phys}$ is generating the equations of motion of the elementary Dirac observables $(\mathcal{O}_b,\mathcal{O}_V,\mathcal{O}_\varphi,\mathcal{O}_{\pi_\varphi})$. At the quantum level this yields the Heisenberg equations for the corresponding operators of these Dirac observables. Switching from the Heisenberg picture to the Schr\"odinger picture we can then find for the dust reference models the respective quantum gravitational Schr\"odinger-like equations  given by 
\bq
\lb{3b11}
i\hbar\frac{\partial}{\partial \tau} \Psi(\ov,\of;\tau)=\left(\widehat{\bf H}^{\rm grav}_\mathrm{phys}+\widehat{\bf  H}^{\varphi}_{\rm phys}\right)\Psi(\ov,\of;\tau).
\eq
Using the action of $\widehat {\bf H}^{\rm grav}_\mathrm{phys}$ and $\widehat{\bf H}^{\varphi}_{\rm phys}$, one finds that
\bqn
\lb{difference}
\widehat{\bf H}_{\rm phys} \Psi( \mathcal O_V,\mathcal O_\varphi;\tau)&=&
\left(\widehat{\bf H}^{\rm grav}_\mathrm{phys}+\widehat{\bf H}^{\varphi}_{\rm phys}\right) \Psi( \mathcal O_V,\mathcal O_\varphi;\tau) \nonumber\\ 
&=&-\alpha  B({\cal O}_V) \partial^2_{\mathcal{O}_\varphi}\Psi(\ov,\of;\tau)-\frac{|\ov|}{2\alpha K}U(\of)\Psi(\ov,\of;\tau)\nb\\
&&~-\alpha C^+( \ov)\Psi({\cal O}_{V + 4},\of;\tau)-\alpha C^0(\ov)\Psi(\ov,\of;\tau)\nonumber \\
&& -\alpha C^-(\ov)\Psi({\cal O}_{V-4},\of;\tau),
\eqn
with coefficients given by 
\bqn
\label{eq:OPHCoeff}
\alpha&=&\frac{1}{2}\left(\frac{6}{8\pi G\gamma}\right)^{3/2}, ~~~
B(\ov)=\left(\frac{3}{2}\right)^{3}K|\ov|||\ov+1|^{1/3}-|\ov-1|^{1/3}|^3,\nb\\
C^+(\ov)&=&\frac{3\pi K G}{8}|\ov+2|||\ov+1|-|\ov+3||, \nb\\
C^-(\ov)&=&C^+(\ov-4), ~~~
C^0(\ov)=-C^+(\ov)-C^-(\ov).
\eqn
The expressions of $C^+$, $C^0$ and $C^-$ are same as in unreduced LQC \cite{aps2006c} albeit with generalization to observables. 
Here as in Dirac quantization of LQC we have expressed the inverse volume in the kinetic energy term of scalar field using Thiemann identity \cite{QSD}, which results in 
\be
\widehat{\mathcal O}_{V^{-1}}  \Psi( \mathcal O_V,\mathcal O_\varphi;\tau) = B({\mathcal O}_V) \Psi( \mathcal O_V,\mathcal O_\varphi;\tau),
\ee
with $B({\mathcal O}_V)$ given by the expression above in (\ref{eq:OPHCoeff}). We thus obtain the evolution equation as a second order quantum difference equation with the same uniform spacing in volume as in conventional LQC which is non-singular. Note that the physical Hamiltonian only relates states which are supported on a `lattice': ${\cal L}_{\pm \epsilon} = \{ \nu = \pm (4n + \epsilon)\}$ with $n \in \mathbb{N}$ and $\epsilon \in (0,4]$. One can then choose a particular `lattice', and as in unreduced LQC, one expects that the action of the physical Hamiltonian and the Dirac observable operators preserve the chosen `lattice'. The coefficients $C^+(\ov)$, $C^0(\ov)$ and $C^-(\ov)$ do not vanish on any chosen `lattice'. Thus, given the wavefunction at volumes ${\cal O}_{V^*+4}$ and ${\cal O}_{V^*}$ for $V^*$ lying in the chosen lattice, the quantum difference equation recursively determines the wavefunction at volume ${\cal O}_{V^*-4}$. The quantum difference equation allows to determine the wavefunction across the to-be classical singularity at zero volume, and in this sense, there is a resolution of big bang singularity at this level. 

As shown in \cite{Kaminski:2008td} the physical Hamiltonian operator of the dust models $\widehat{\bf H}_{\rm phys}$ is self-adjoint in ${\cal H}_{\rm phys}={\cal H}_{\rm phys,grav}\otimes{\cal H}_{\rm phys,\varphi}$ which is the physical Hilbert space in the dust models. Note that in \cite{Kaminski:2008td} only the case of a massless scalar field with a cosmological constant was considered, however as discussed in \cite{hp2011} the property of self-adjointness is preserved for any non-exotic matter contribution added to the gravitational sector. One can include potentials in this setting which are equivalent to adding different cosmological constants for different time slices. Using  above results, self-adjointness of the physical Hamiltonian follows.  Further, also the elementary Dirac observables $\widehat{\cal O}_V,\widehat{\cal O}_b,\widehat{\cal O}_\varphi,\widehat{\cal O}_{\pi_\varphi}$ are self-adjoint operators in ${\cal H}_{\rm phys}$ and the physical Hamiltonian $\widehat{\bf H}_{\rm phys}$ can then be understood as a function of these elementary operators.

Thus using the dust reference fields as clocks we have found the corresponding quantum gravitational Schr\"odinger-like equation  which is a difference equation with a uniform spacing in four Planck volumes. The latter is a primary characteristic of the LQC quantum difference equation in the $\bar \mu$-scheme which results in a successful resolution of the big bang singularity replacing it with a quantum bounce \cite{aps2006c,acs2010}.
In contrast to the unreduced LQC model \cite{aps2006c}, in the dust models we can without further technical complications include a potential for the inflaton since even then the resulting physical Hamiltonian operator stays time-independent. As mentioned before in the model of type II \cite{hp2011,hp2012} this would be also possible but up to now an analysis of the physical properties of such a model is not available in the literature. 

An important next step will be  to investigate solutions of the above Schr\"odinger equation. Compared to the unreduced LQC model this is technically more involved  since we have one more degree of freedom at the level of the physical Hilbert space. As a result, numerical simulations with the full quantum difference equation will be more demanding. Nevertheless, thanks to recent developments in including supercomputing efforts in LQC \cite{ps2018} and efficient algorithms \cite{chimera}, the situation above is expected to be no more than computational requirements of analyzing difference equations in anisotropic models which have been successfully analyzed recently \cite{djms2017}. We expect that as in conventional LQC, numerical simulations with quantum difference equations would show that singularity resolution occurs via a quantum bounce in the Planck regime. Note that since we use the usual kinematical LQC representation for ${\cal H}_{\rm phys}$ here the physical Hilbert space on which the Schr\"odinger-like equation is formulated is still non-separable. However, given the structure of the quantum difference equation discussed above regarding `lattices' ${\cal L}_{\pm \epsilon}$, we expect that if we restrict on solutions of the quantum dynamics there is a super-selection and a Hilbert space associated to the solutions can become separable. 
%%%%%%%%%%%%%%%%
\subsection{The physical Hilbert space and Hamiltonian in the Klein-Gordon scalar field model}
After discussing the quantum dynamics of the two dust models in detail in the last subsection we will be  brief in the presentation of the Klein-Gordon scalar field model where many of the steps performed in the dust models can be carried over. We consider the same ${\cal H}_{\rm phys,\varphi}$ for the inflaton but slightly modify the inner product for the gravitational part and introduce as in \cite{aps2006c} the following inner product:
\begin{equation}
{\cal H}_{\rm phys,grav}^{\rm B}:=\overline{\rm{span}(|\nu\rangle\, :\, \nu\in\mathbb{R})}\quad{\rm with}\quad 
\langle \nu \, ,\, \nu'\rangle_B:=\langle \nu \, ,\, B(\mathcal{O}_V)\nu'\rangle,
\end{equation}
where $B(\mathcal{O}_V)$ is given in (\ref{eq:OPHCoeff}). The reason for this modification is that we want the physical Hamiltonian of the scalar field model to be self-adjoint. The elementary operators acting on ${\cal H}_{\rm phys}$ act in a similar way as in the dust model shown in (\ref{eq:ActionOpgrav}) and (\ref{eq:ActionOpphi}) respectively.  As before we can go from the Heisenberg picture that one obtains after the reduced phase space quantization to the Schr\"odinger picture. 
In case of the Klein-Gordon scalar field clock, the Schr\"odinger-like equation  is given by 
\bq
\lb{KG-sch}
i\hbar\frac{\partial}{\partial \tau_\chi} \Psi(\ov,\of;\tau)=\sqrt{|-2\hat{B}^{-1}({\mathcal O}_V)(\widehat{\bf H}^{\rm grav}_\mathrm{phys}+\widehat{\bf H}^{\varphi}_{\rm phys})|} ~~ \Psi(\ov,\of;\tau),
\eq
with the action of $(\widehat {\bf H}^{\rm grav}_\mathrm{phys}+\widehat{\bf H}^{\varphi}_{\rm phys}))\Psi(\ov,\of;\tau)$ given by  (\ref{difference}).  Here the factor of 
$B({\mathcal O}_V)^{-1}$ arises from the inverse volume term in the Klein-Gordon clock and we included in absolute value for the expression under the square root.

The operator under the square root inside the absolute value, up to the factor 2, can  be identified with the $\hat{\Theta}$ operator in \cite{aps2006c} written in terms of Dirac observables and generalized to the case of an inflationary potential. As discussed in the case of the dust model, and shown in \cite{Kaminski:2008td} this is a self-adjoint operator on the physical Hilbert space ${\cal H}_{\rm phys}^{\rm B}$. In this case, unlike the dust models, the spectrum of the physical Hamiltonian is expected to be more non-trivial capturing some features discussed earlier for a positive cosmological constant with a massless scalar field as a clock \cite{ds1,*ds2} but now with an additional complexity of a potential which corresponds to choosing different values of a positive cosmological constant at different times $\tau_\chi$. 
Further, unlike the dust clock, we now have a square root operator arising from the physical Hamiltonian in this case. 
Hence, the structure of the Schr\"odinger equation of the scalar field model appears more complicated compared to the dust model case at this stage. We found above that the evolution equation in the reduced phase quantization turns out to be a quantum difference equation with a quantum discreteness set by the underlying quantum geometry via the minimum eigenvalue of the area operator in LQG.  This is the same structure as in conventional LQC which results in a singularity resolution.
%%%%%%%%%%%%%%%%%%
\subsection{Effective dynamical equations for the dust clocks}
As in the Dirac quantization of LQC, insights in to the physical evolution can be obtained by using an effective spacetime description which captures the quantum evolution in LQC  for different models quite accurately \cite{aps2006b,dgs2014,djms2017,ps2018}. This effective description allows obtaining numerical solutions on a differentiable spacetime which encodes quantum gravitational effects. This effective spacetime description is obtained from a geometrical formulation of the quantum theory using an appropriate choice of sharply peaked states which  results in an effective Hamiltonian constraint \cite{vt}. Extensive numerical simulations carried out with a wide variety of states using quantum difference equations show that the dynamics obtained from the effective Hamiltonian is an excellent approximation to quantum dynamics for states which bounce at volumes much larger than the Planck volume \cite{dgs2014,djms2017,ps2018}. For such states the difference between the expectation values of the relevant physical observables and their counterparts in the effective dynamics turns out to be negligible. On the other hand for states which are widely spread or bounce at volume close to Planck volume, there can be departures between the quantum and effective dynamics \cite{dgms2014}. In this paper we would assume that suitable sharply peaked states exist in the reduced phase space quantization which lead to a reliable effective spacetime description for different reference fields. 
 In this effective dynamics setting we first present dynamical equations for dust clocks, which is followed by those for Klein-Gordon scalar field clock. As in the usual case in LQC, we work in the approximation where effects from inverse volume operators which are significant only near the Planck volume can be ignored. In the solutions discussed below, the volume of the universe remains much larger than the Planck volume even at bounce which is consistent with above approximation. Further, we will choose the orientation of the triad to be positive.

For the dust clock, following the procedure used in Dirac quantization in LQC, an effective physical  Hamiltonian written in terms of Dirac observables ${\cal O}_V$, 
${\cal O}_b$, ${\cal O}_\varphi$, and ${\cal O}_{\pi_\varphi}$, can be written as: 
\bq
\lb{hamiltonian}
{\bf H}^\mathrm{FLRW, dust}_\mathrm{eff}=-\frac{3 \ov}{8\pi  G \lambda^2\gamma^2}\sin^2(\lambda \ob)+\frac{\opf^2}{2 \ov}+ \ov U(\of) ,
\eq
where $\ob$ is defined as  $\ob=\mathcal O_c/{\cal O}_{|p|}^{1/2}$ and $\ov=\mathrm{sgn}(\mathcal O_p){\cal O}_{|p|}^{3/2}$ and $\lambda = \sqrt{\Delta}$. One can then use $\ob$ and $\ov$ as phase space variables, which satisfy 
$\{\ob, \ov\}=4\pi G \gamma$. The resulting Hamilton's equations in the reduced phase space are
\bqn
\lb{ham1}
\frac{\der {\ov}}{\der \tau} &=&\frac{3 \ov}{2\lambda \gamma}\sin(2\lambda \ob), \quad \frac{\der {\ob}}{\der \tau}=-\frac{3\sin^2(\lambda \ob)}{2\gamma\lambda^2}-4\pi G \gamma {\mathcal O}_{P_\varphi},\\
%\lb{ham3}
\lb{ham4}
\frac{\der {\of}}{\der \tau}  &=& \frac{\opf}{\ov},~~~~~~~~~~~~~~~ \frac{\der {\opf}}{\der \tau} =-\ov \frac{\der U(\of)}{\der \of} .
%\lb{ham2}
\eqn
Here $\mathcal{O}_{P_\varphi}$ denotes the observable corresponding to the pressure of the inflaton field $\varphi$
\bq
\lb{3b6}
\mathcal{O}_{P_\varphi}=\frac{\opf^2}{2{\mathcal{O}}_V^2}-U(\of). 
\eq
Since the dust clocks are pressure-less they do not contribute to total pressure which is captured by the above observable.  
In contrast, the total energy density observable also includes contribution from the dust:
\bq
\lb{3b7}
\mathcal{O}_\rho= \frac{\opf^2}{2{\mathcal{O}}_V^2}+U(\of) + \frac{ {\mathcal E}^\mathrm{dust}}{\ov} ~,
\eq
where ${\mathcal E}^\mathrm{dust}$ is defined as the negative physical Hamiltonian given by (\ref{hamiltonian}), and the ratio  ${\mathcal E}^\mathrm{dust}/\ov$   corresponds to the energy density of the dust clock.  Note that in the Brown-Kucha\v r case, the energy density for dust can also take negative values. 

For understanding cosmological dynamics it is useful to derive the modified Friedmann equation which can be obtained from \eqref{ham1} by computing the square of the observable for the Hubble rate $H = \dot V/ 3 V$, which using \eqref{hamiltonian} yields
\bq
\lb{friedmann}
\mathcal{O}_H^2 = \frac{\dot{\mathcal{O}}_V^2}{9 \ov^2} = \frac{8\pi G}{3}\mathcal{O}_\rho\left(1-\frac{\mathcal{O}_\rho}{\rho_{\mathrm{max}}}\right).
\eq
Here $\rho_{\mathrm{max}}$ is a constant determined by the area gap,
\bq
\lb{maxdensity}
\rho_{\mathrm{max}}=\frac{3}{8\pi G\gamma^2\lambda^2}.
\eq
When $\mathcal{O}_\rho = \rho_{\mathrm{max}} \approx 0.41 \rho_{\mathrm{Pl}}$, the observable corresponding to the Hubble rate vanishes and the Dirac observable of the volume $\ov$ reaches its minimum value when the bounce occurs. At this value the energy density is bounded above by $\rho_{\mathrm{max}}$. In the limit where $\lambda \rightarrow 0$, $\rho_{\mathrm{max}} \rightarrow \infty$ and the observable $\mathcal{O}_\rho$ is unbounded from above. In this limit, the big bang singularity is recovered. 

As compared with the modified Friedmann equation obtained in an effective description of the Dirac quantization of LQC, equation (\ref{friedmann}) has the same form and the same maximum energy density at which the bounce takes place. The only difference lies in the composition of the energy density  which is now made up of two ingredients, the inflaton and the contribution from dust clock. It is also straightforward to show that the form of the modified Raychaudhuri equation in the relational formalism is the same as its counterpart in LQC. This is owing to the property of the observable map in (\ref{3b3}) which ensures all the dynamical equations can be lifted to the observable level in a straightforward way once the the contribution from the dust field to the total energy density is taken into account. 
 
Finally, the Klein-Gordon equation of the scalar fields can be derived from the Hamilton's equations in a straightforward way. Taking the time derivative of (\ref{ham4}) and combining it with (\ref{ham1}), one finds,
\bq
\lb{kg}
{\ddot{\mathcal O}}_\varphi +3 \mathcal{O}_H {\dot{\mathcal O}}_\varphi+U_{,\of}=0.
\eq
Using this equation it is straightforward to show that the continuity equation is satisfied for the observables of energy density and pressure of the scalar field. Further, the observable for the total energy density which includes contribution from dust clocks also satisfies the continuity equation. 

In the next section, we use these dynamical equations to understand some features of the resulting non-singular dynamics with dust clocks. 

\subsection{Effective dynamical equations for the Klein-Gordon scalar field clock}
                           
Similar to the dust model, we assume that an effective spacetime description also exists in the Klein-Gordon scalar field clock model. By using the $\mathcal O_b$, $\mathcal O_V$ variables and the same substitutions in the classical physical Hamiltonian (\ref{phyhamScal}), the effective Hamiltonian in the Klein-Gordon scalar field model can be shown to be
\bq
{\bf H}^\mathrm{FLRW, scalar}_\mathrm{eff}=\sqrt{-2\ov\left(-\frac{3 \ov}{8\pi  G \lambda^2\gamma^2}\sin^2(\lambda \ob)+\frac{\opf^2}{2 \ov}+ \ov U(\of)\right)},
\eq
It is then straightforward to derive Hamilton's equations which read
\bqn
\lb{h1}
\frac{\der {\ov}}{\der \tau_\chi} &=&\frac{3 \mathcal O_{N} \ov}{2\lambda \gamma}\sin(2\lambda \ob),~~~ \frac{\der {\ob}}{\der \tau_\chi}=-4\pi G \gamma  \mathcal O_{N} \left({\mathcal O^\mathrm{scalar}_\rho+\mathcal O}^\mathrm{scalar}_{P}\right), \\
%\lb{h2}
\lb{h3}
\frac{\der {\of}}{\der \tau_\chi}&=&\frac{ \mathcal O_{N}\opf}{\ov},~~~~~~~~~~~~~~~ \frac{\der {\opf}}{\der \tau_\chi}=- \mathcal O_{N}\ov \frac{\der U(\of)}{\der \of},
%\lb{h4}
\eqn
where  $\mathcal O_{N}=-\ov/{\bf H}^\mathrm{FLRW, scalar}_\mathrm{eff}$ and  the total energy density and the pressure are given by
\bqn
\lb{scalardensity}
\mathcal O^\mathrm{scalar}_\rho&=&\frac{\opf^2}{2{\mathcal{O}}_V^2}+U(\of) + \frac{ \mathcal E^2_\chi }{2\ov^2} ~,\\
\mathcal O^\mathrm{scalar}_P&=&\frac{\opf^2}{2{\mathcal{O}}_V^2}-U(\of) + \frac{ \mathcal E^2_\chi }{2\ov^2} ~.
\eqn
In the above formulas,  $\mathcal E_\chi=-{\bf H}^\mathrm{FLRW, scalar}_\mathrm{eff}$ is a constant of motion. It should be noted that in the Hamilton's equations all the Dirac observables  are evolved with respect to the Klein-Gordon scalar physical time $\tau_\chi$. This physical time is related to the proper time $\tau$ (which is also  physical dust time) by $\frac{\der \tau}{\der \tau_\chi}=\mathcal O_{N}$. Further, given that the lapse function is negative, $\tau_\chi$ is negative of the scalar field clock  used in LQC \cite{aps2006c}. 

The forms of the Friedmann and Raychaudhuri equations when expressed in Klein-Gordon scalar physical time  $\tau_\chi$ are slightly different than the ones in  proper time $\tau$. For example, with respect to the Klein-Gordon scalar field clock, the Friedmann equation in effective description takes the form 
\bq
\tilde H^2=\mathcal O^2_N \frac{8\pi G}{3}\mathcal O_\rho\left(1-\frac{\mathcal O_\rho}{\rho_\mathrm{max}}\right),
\eq
where we have defined the Hubble rate in the Klein-Gordon scalar field model as $\tilde H=\der \mathcal O_V/\der \tau_\chi$. It is important to note that although the above equation is not same as \eqref{friedmann} because of different choice of clocks,  they have the same physical content. One can transform from $\tau_\chi$ to $\tau$ in above equation to obtain \eqref{friedmann}. As in GR, the physical predictions are by no means dependent on the specific time coordinates used in the dynamical equations.  In particular, the above equation still yields the bounce at the maximum energy density  given by (\ref{maxdensity}) and the super-inflationary phase, defined as the regime in which the time derivative of the Hubble rate is positive,  occurs in the range  $\rho_\mathrm{max}/2 \le \mathcal O^\mathrm{scalar}_\rho \le \rho_\mathrm{max}$. 

%One can also study the evolution in $\tau$ which is a monotonic function of  $\tau_\chi$. Then the dynamical equations of the fundamental observables  take the same form as those in the dust models, but, the difference lies in the expression of the energy density and pressure which carry the fingerprints of the different types of clocks. Also with respect to the proper time $\tau$, the modified Friedmann and Raychaudhuri equations take the same forms as those in LQC. Further, the Klein-Gordon equation of the massive scalar field $\mathcal{O}_\varphi$ will take the same form as (\ref{kg}) in the physical dust time $\tau$. 

Finally to summarize, in this section, using the techniques from LQC, we have quantized the reduced phase space of a  spatially flat FLRW universe with both dust clocks and a massless Klein-Gordon scalar clock. We have derived the Schr\"{o}dinger-like quantum difference equations and the Hamilton's equations of the effective dynamics for the three different types of clocks. These effective equations lay the basis for the numerical analysis of the background dynamics of the spatially flat FLRW universe in the reduced phase space framework which is discussed in the following.

\section{Numerical analysis of the effective dynamics with  the dust  and massless Klein-Gordon scalar field clocks}
\label{sec:numeric}
\renewcommand{\theequation}{4.\arabic{equation}}\setcounter{equation}{0}

In this section, starting from the Hamilton's equations obtained in an effective spacetime description of the reduced phase quantization of a flat FLRW universe, numerical solutions of the background dynamics with respect to the reference fields are found when an inflaton field is present. Both the references fields and the inflaton field  are minimally coupled to gravity.  We discuss two separate cases: one for dust clocks as reference fields and another one with a Klein-Gordon scalar reference field. The inflationary potential is chosen to be the Starobinsky potential
\bq
\lb{star}
U=\frac{3m^2}{16\pi G}\left(1-e^{-\sqrt{\frac{16\pi G}{3}}\of}\right)^2,
\eq
where the mass is fixed to be $m=2.44\times10^{-6}$ from the observational data  \cite{lsw2019}. 
The initial conditions are chosen at the bounce where the total energy density reaches its maximum value. 
A property of the physical Hamiltonian in this reduced phase quantization is that at the bounce $\ob$ equals $\pi/2\lambda$. This leaves the Dirac observables $\mathcal O_V$, $\mathcal O_\varphi$ and $\mathcal O_{\pi_\varphi}$  to be specified for the initial conditions. For our numerical solutions, we set the Dirac observable of the volume at the bounce to $10^3$ in Planck units. Then the choice of the initial values of $\mathcal O_\varphi$ and $\mathcal O_{\pi_\varphi}$  determines the value of the physical Hamiltonian at the bounce, or equivalently, the initial value of ${\cal E}^{\mathrm{dust}}$ in the dust model and $\mathcal E_\chi$ in the Klein-Gordon scalar field model. Apart from studying singularity resolution with different clocks, we will be also interested in understanding the way energy density of the different clocks may have any imprint on the cosmological dynamics. For this reason, we parametrize the space of the initial conditions by $(\mathcal O_{\varphi},\mathcal O_{\rho^\mathrm{clock}})$. 

The effects of the dust reference fields on the background dynamics in the classical theory for both Gaussian and Brown-Kucha\v r dust models were discussed recently by the authors in \cite{ghls2020}. There it was found that changing the initial dust energy density can effectively change the inflationary e-foldings. The numerical analysis of the classical theory in \cite{ghls2020} started with initial conditions when the inflaton starts from the left wing of the Starobinsky potential as in this manuscript. It was found thatwhen the magnitude of the initial negative dust energy exceeds an upper bound determined by the initial conditions for the other physical quantities, the inflationary phase is replaced by a recollapse of the universe which finally results in a big crunch singularity. In classical cosmology,  solutions with either positive or negative dust energy density are past incomplete with an inevitable big bang singularity when evolved backward in time. In the following, we will show that all  singularities encountered in the classical theory in \cite{ghls2020} are resolved and replaced by a quantum bounce when the classical spatially flat FLRW spacetime is loop quantized as in the last section. In all the representative cases discussed below, we focus on  the situation when the inflaton initially rolls down the left wing of the Starobinsky potential. 
Since the energy density of the dust field can also take negative values in the Brown-Kucha\v r dust model, we consider both positive and negative dust energy densities in our analysis. Note that for the background dynamics, the Gaussian and Brown-Kucha\v r dust clocks yield no difference when the energy density is positive. For this reason, in the following, cases for positive dust energy density we refer to both of the dust clocks. Further, in the discussion below all values of the Dirac observables are considered in Planck units.

\begin{figure}
{
\includegraphics[width=8cm]{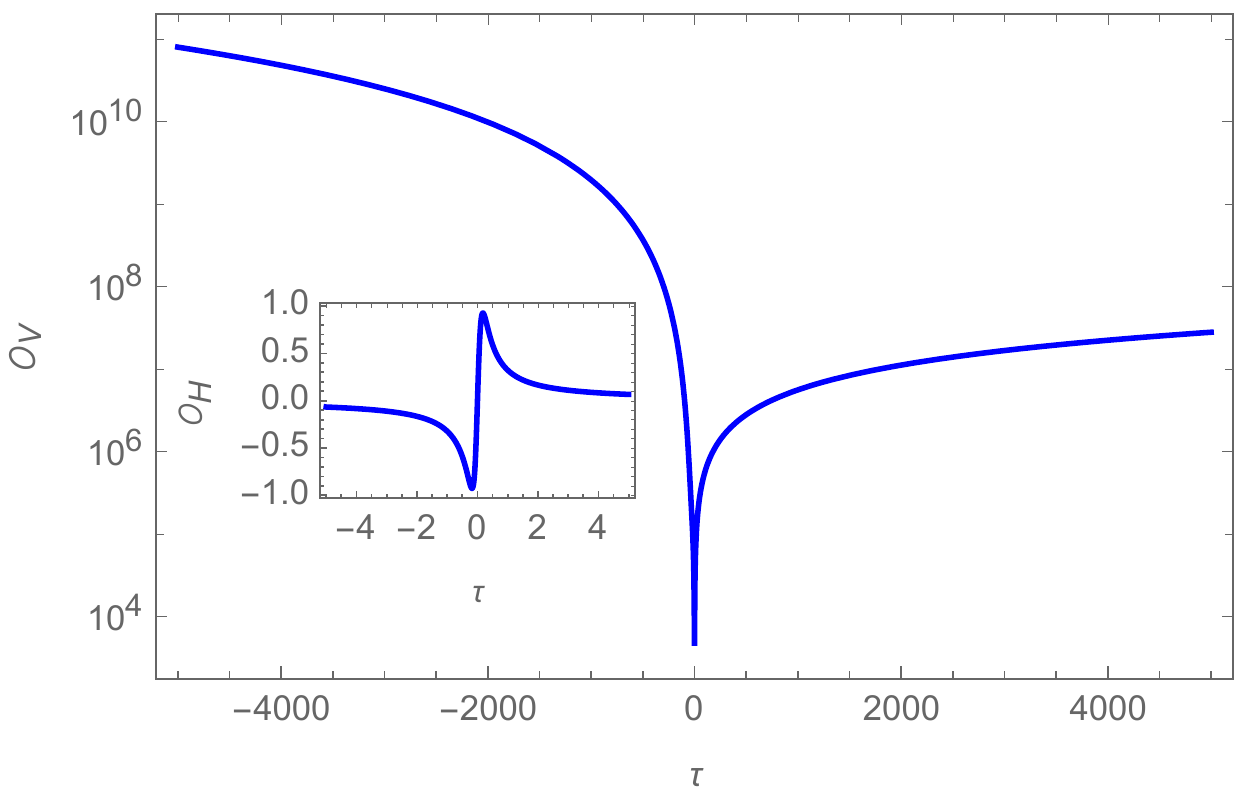}
\includegraphics[width=7.6cm]{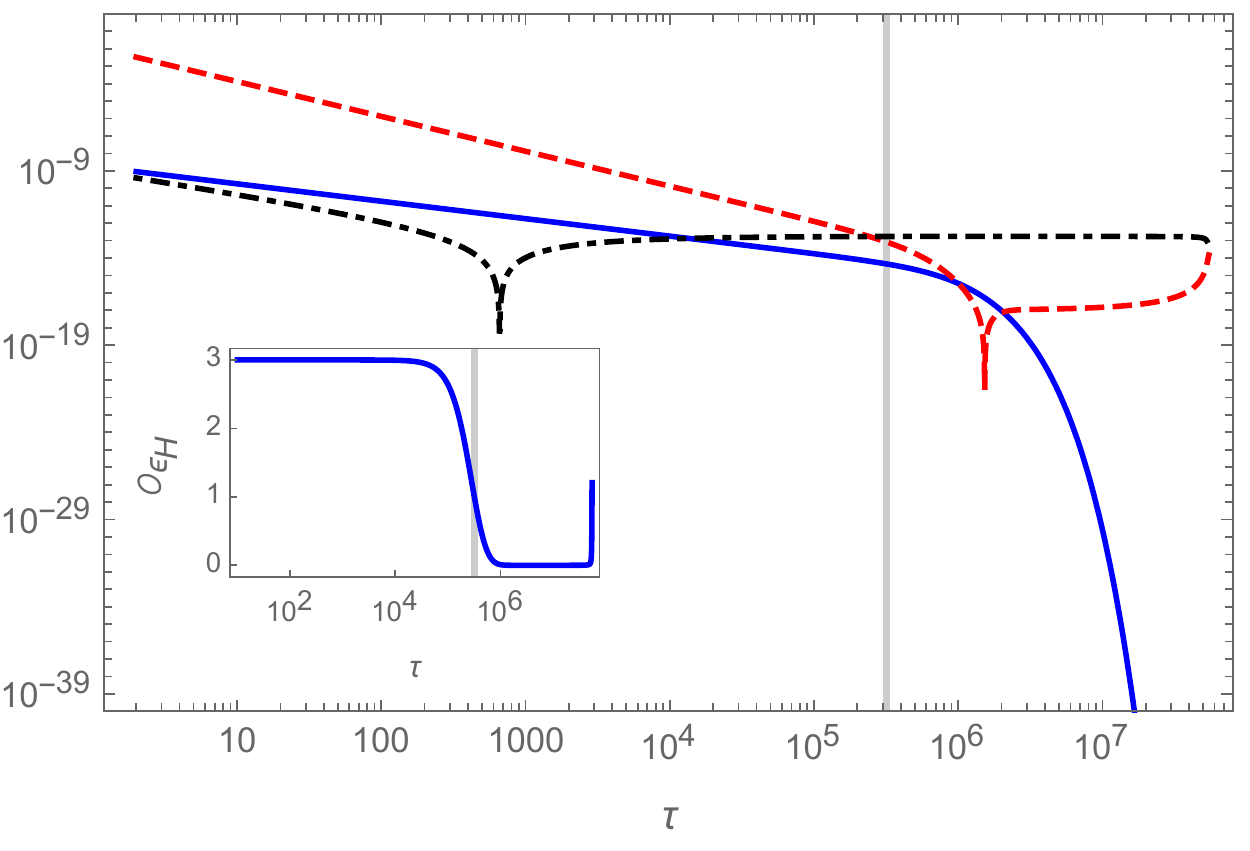}
}
\caption{In this figure, with the initial conditions given in  (\ref{initial1}), the evolution of the volume Dirac observable near the bounce is depicted in the left panel where the inset figure shows the behavior of the Hubble rate across the bounce. In the right panel, we compare the time evolution of the kinetic (red dashed) and potential (black dot-dashed) energies  of the scalar field with the dust energy density (blue solid). The inset plot of the right panel depicts the time evolution of the first slow-roll parameter till the end of inflation. The vertical lines in these graphs indicate the onset of inflation at $\tau=3.17\times10^5$.} 
\label{f1}
\end{figure}

\subsection{Dust clocks with a positive energy density}
In this subsection we discuss a representative case of initial conditions chosen for dust clocks with a positive energy density.  As mentioned earlier, for numerical analysis  the initial conditions for geometric variables are ${\cal O}_{b_i} =\pi/2\lambda$ and ${O}_{V_i} = 1000$.  Initial conditions for other variables are chosen as 
\bq
\lb{initial1}
\mathcal O_{\varphi_i}=-1.45,\quad \quad \mathcal O_{\rho^\mathrm{dust}_i}=10^{-8},
\eq
where the subscript `$i$' stands for the initial values of the relevant quantities at the bounce. With these initial values, we find $\mathcal O_{\pi_{\varphi_i}}=900$. Note that for the  above initial conditions, $\mathcal O_{\pi_{\varphi_i}}$ can be either positive or negative, but the latter does not yield a viable phase of inflation and hence is not considered.\footnote{When the initial velocity is negative, only the positive values of the inflaton field at the initial time can lead to successful inflation later on. }  Also, we choose these particular values for $\mathcal O_{\varphi_i}$ and $\mathcal O_{\pi_{\varphi_i}}$ since they yield slightly greater inflationary e-foldings than 60 in LQC in the absence of dust reference fields. In particular, these initial conditions without any contribution from dust clocks yield 63.9 inflationary e-foldings (for details on number of e-foldings for various initial conditions in LQC for Starobinsky potential see \cite{lsw2019}). One of our goals in the following will be  to understand the effect of dust clocks on the number of e-foldings even when the energy density of clocks is much smaller than the inflaton energy density.

Fig. \ref{f1} shows the results for above initial conditions. 
In the left panel, we see that a non-singular bounce takes place at $\tau=0$ with the inset showing the behavior of the Dirac observable for the Hubble rate.  The super-inflationary phase starts from the bounce and ends at the moment when the Hubble rate reaches its maximum at around $\tau=0.18$,  yielding a total of $0.12$ e-foldings.  The universe at the bounce is  dominated by the kinetic energy of the scalar field as shown in the right panel of  the figure. As the inflaton first rolls down the left wing and then climbs up the right wing of the potential, it slows down with a positive velocity before the turnaround point. The slow-roll inflation  takes place shortly after the turnaround point and ends when the inflaton reaches the bottom of the potential. Throughout the period from the bounce to the end of the inflation, the dust field  plays a subdominant role as its energy density stays much less than the energy density of the scalar field. This behavior of the dust clocks becomes  more obvious in the slow-roll inflationary phase in which  the dust energy density  exponentially  decreases while the energy  density of the scalar field remains almost constant. The presence of the inflationary phase can be further confirmed by the inset plot of the observable for the slow-roll parameter $\mathcal O_{\epsilon_H}$ defined as 
\bq
\lb{slowroll}
\mathcal O_{\epsilon_H}=\frac{4\pi G}{\mathcal O_H^2}\left(\mathcal O_{\rho_\varphi}+\mathcal O_{\pi_\varphi}+\frac{\mathcal E^\mathrm{dust}}{\mathcal O_V}\right).
\eq
Inflation takes place at $\mathcal O_{\epsilon_H}=1$  marked by the gray vertical line in the figure and ends when $\mathcal O_{\epsilon_H}$ grows up to  unity again after the exponential expansion of the universe,  yielding a total of $63.1$ inflationary e-foldings. 
%%%%%%%%%
%%%%%%%%%%
\subsection{Dust clocks with a negative energy density}
In the Brown-Kucha\v r model, the energy density of the dust can be negative. If the energy density of such a phantom dust is sufficiently large, the universe undergoes a recollapse before inflation can start. Classically such a universe encounters a big crunch singularity \cite{ghls2020}, which is avoided by loop quantum effects, and such initial conditions result in several cycles before inflation can start. Note that such a negative phantom dust field is not consistent with the well-known $\Lambda$CDM model of the current universe which is in excellent agreement with the latest observations since the latter  only allows for a positive cosmological constant.   Nevertheless, we discuss this case to show how different choices of initial conditions for the dust clock can result in very different dynamics.  
In the following,  we first discuss a case when the energy density of dust allows inflationary dynamics to occur without any such cycles. This is followed by examples where the energy density of the dust clock is so negative that the universe undergoes cycles of contraction and expansion before inflation can set in. 

\begin{figure}
{
\includegraphics[width=8.1cm]{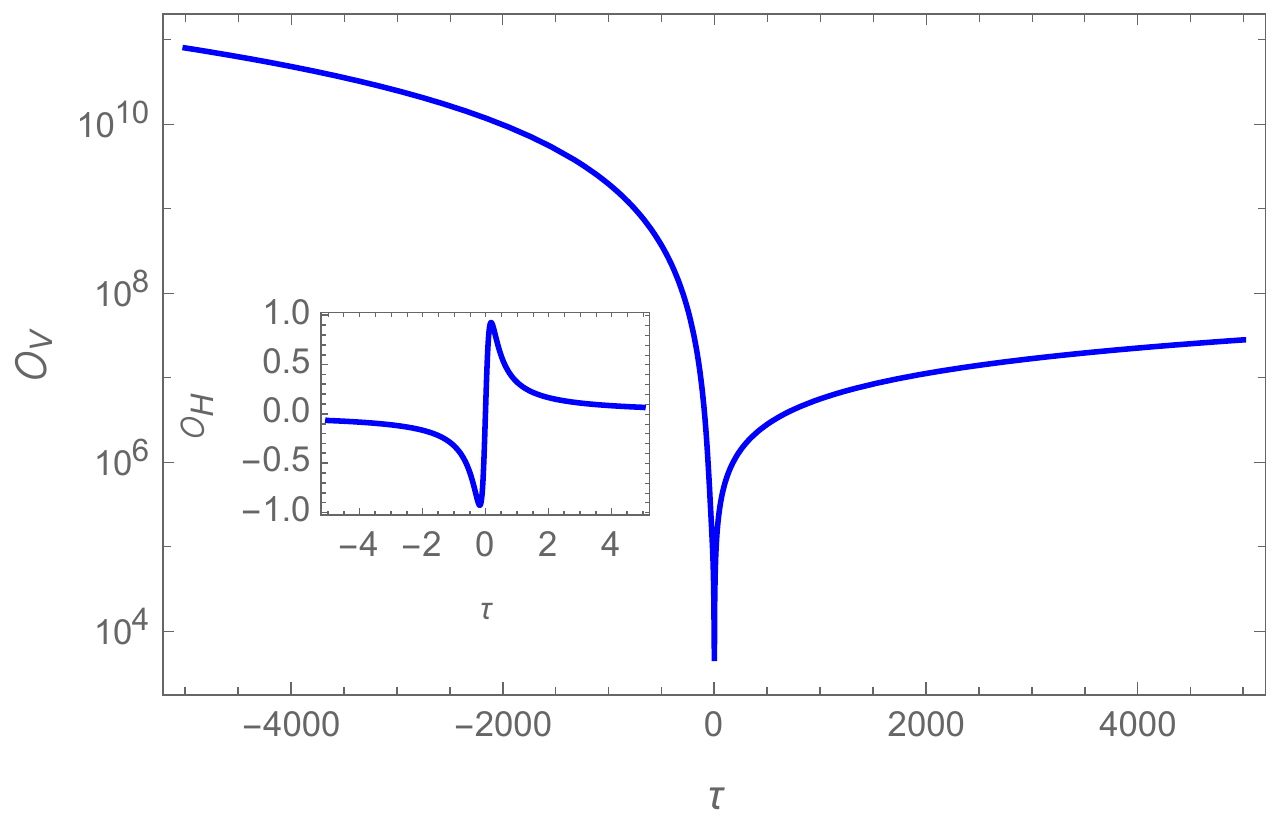}
\includegraphics[width=7.6cm]{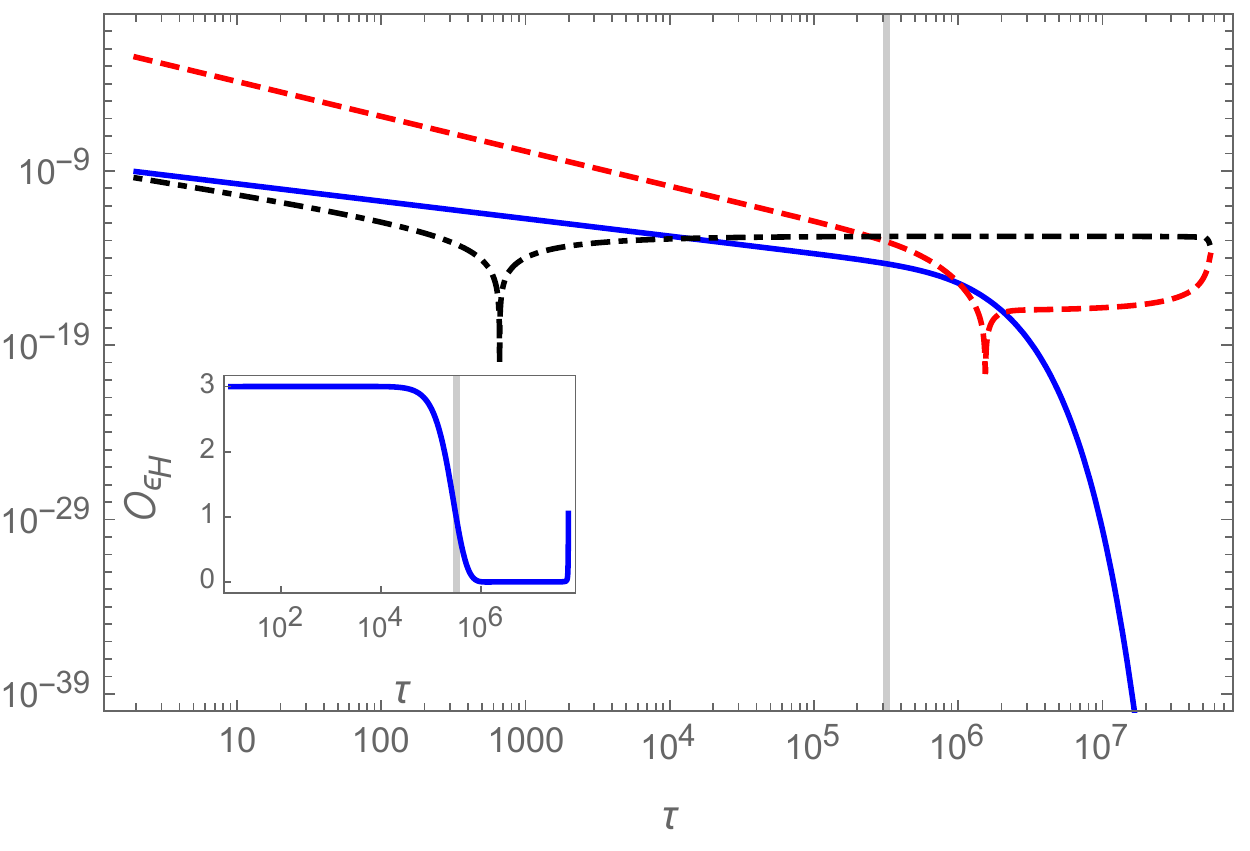}
}
\caption{These plots correspond to the initial conditions given in (\ref{initial2}) for a negative energy density of the dust clock, the evolution of the volume Dirac observable and the Hubble rate is depicted near the bounce in the  left panel. The right panel shows the time evolution of the kinetic and potential energies of the scalar field and the first slow-roll parameter (in the inset plot) where the gray vertical lines indicate the onset of inflation at $\tau=3.18\times10^5$.  Inflation  ends at  $\tau= 5.60\times10^7$,  yielding a total of $64.7$ inflationary e-foldings.}
\label{f3}
\end{figure}

The first set of initial conditions for negative energy density are chosen to differ from conditions in (\ref{initial1}) just by a negative sign of the dust energy density,
\bq
\lb{initial2}
\mathcal O_{\varphi_i}=-1.45, \quad \quad \mathcal O_{\rho^\mathrm{dust}_i}=-10^{-8},
\eq
which also set $\mathcal O_{\pi_{\varphi_i}}=900$. As before we choose ${\cal O}_{b_i} =\pi/2\lambda$ and ${O}_{V_i} = 1000$ in Planck units. 
In this case, the qualitative evolution of the universe is similar to the case with the positive dust energy density discussed in the last subsection. As shown in the left panel of Fig. \ref{f3},   the Dirac observables of the volume  and the Hubble rate evolve continuously across the bounce at $\tau=0$.  Also, the dust field is subdominant at all times.  The difference from Fig. \ref{f1} lies in the exact number of e-foldings  of the inflationary phase, which in this case is $64.7$, larger than the inflationary e-foldings in Fig. \ref{f1}. This increase in the number of e-foldings is tied to the higher value of $\mathcal O_\varphi$ at the onset of inflation when dust energy density is negative, which occurs because of the decrease in the Hubble friction. An increase in the magnitude of the negative dust energy density further  decreases the Hubble friction which makes the inflaton turn around at a higher value of $\mathcal O_\varphi$ in the Starobinsky potential.

\begin{figure}
{
\includegraphics[width=8cm]{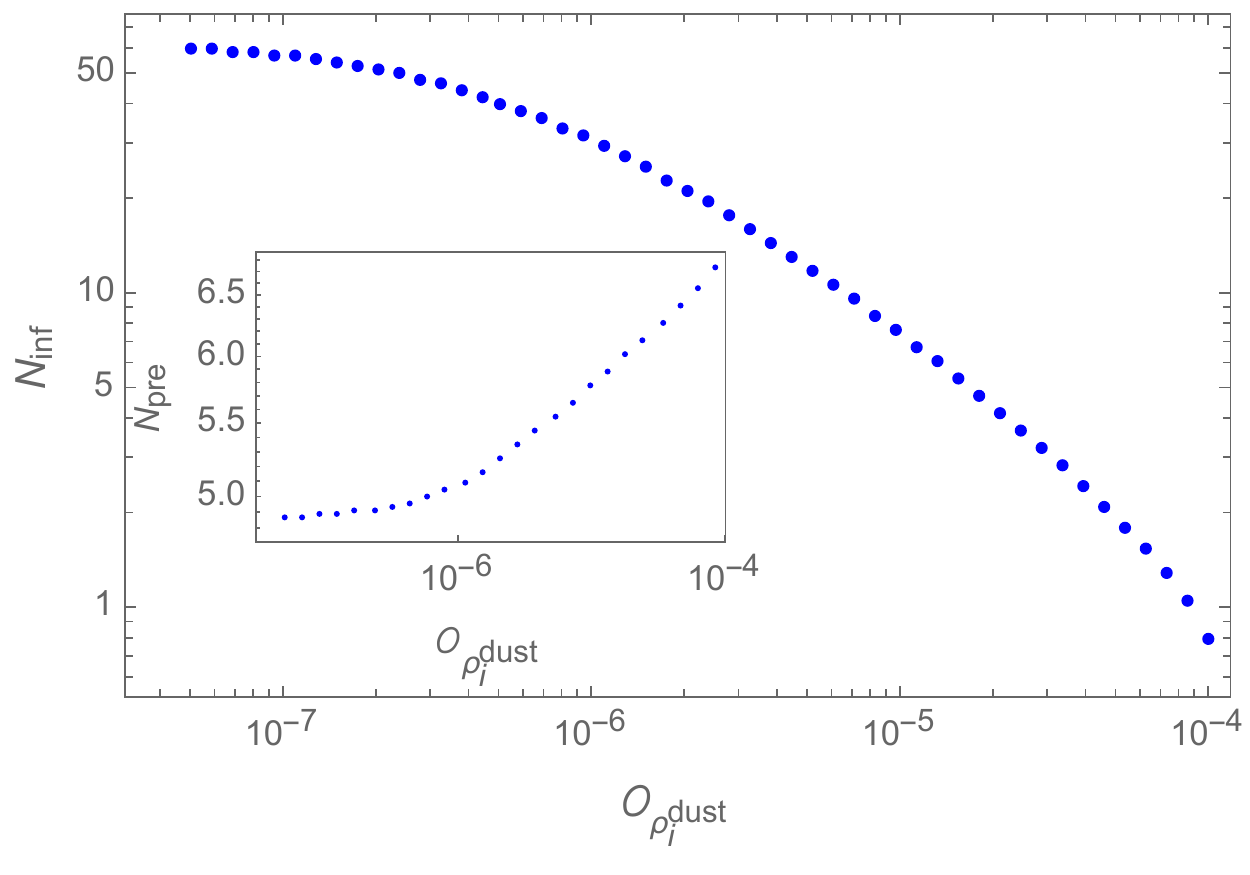}
\includegraphics[width=8cm]{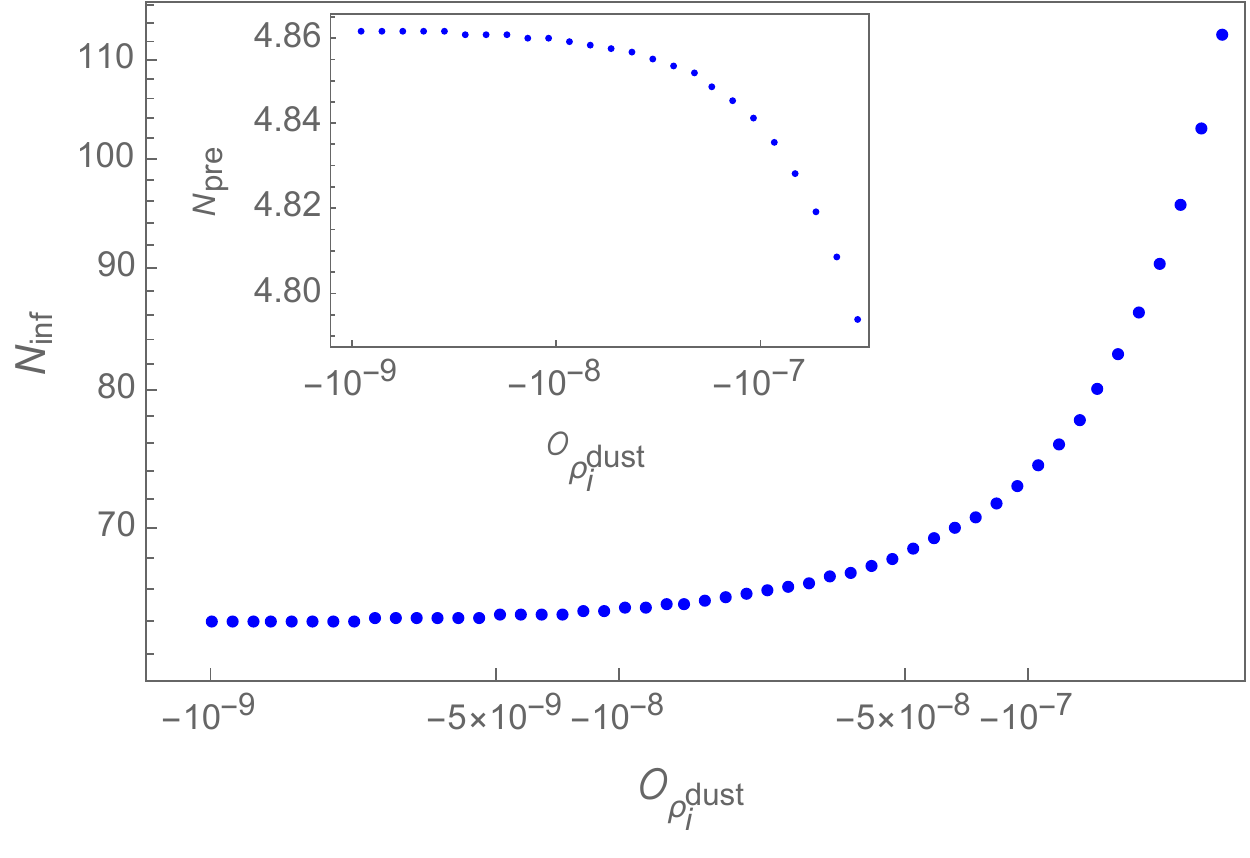}
}
\caption{With the initial conditions set at the bounce and  $\mathcal O_{\varphi_i}=-1.45$,  we compare the effects of the positive (left panel) and negative (right panel) dust energy densities on the e-foldings of the the pre-inflationary (in the inset subfigures) and the inflationary phases. }
\label{f4}
\end{figure}

The effects of the dust fields on the e-foldings of  the preinflationary and inflationary phases are depicted in Fig. \ref{f4} when the initial value of the scalar field is fixed to be $-1.45$ at the bounce for both positive and negative dust energy densities. In the left panel,  the initial positive dust energy density  increases in small increments from $5\times10^{-8}$ to $10^{-4}$ which constitutes $0.000012\%$ to $0.024\%$ of the total energy density at the bounce.  It shows that the  e-foldings of  the inflationary (pre-inflationary) phase decreases (increases) with an increasing positive dust energy density at the bounce.\footnote{ Note that the pre-inflationary e-foldings are counted from the bounce to the onset of inflation, and includes the e-foldings from super-inflation.} This is because the Hubble friction becomes larger when the initial positive dust energy density is increased resulting in a smaller value of  the scalar field at the onset of inflation. In the extreme  case, when the initial positive dust energy is so large  that the inflationary phase  can disappear. Such extreme values clearly do not correspond to dust acting as  a good clock.  As a result, for a given value of $\mathcal O_{\varphi_i}$,  there exists an upper bound on the value of initial dust density for which the inflationary phase does not occur. For   $\mathcal O_{\varphi_i}=-1.45$, this bound turns out to be $1.38\times 10^{-4}$. In the right panel of  Fig. \ref{f4}, we find  the negative dust energy has an opposite effect on the number of  the preinflationary  and inflationary  e-foldings as compared with the positive dust energy. But similar to the latter case, for any given initial values of the scalar field, there is  an upper bound $\mathcal O_{\rho_\mathrm{upper}}$ (defined positive) on the  magnitude of the negative dust energy density to ensure the dynamics of the LQC universe is not qualitatively altered by the dust reference field. When $|\mathcal O_{\rho^\mathrm{dust}_i}|<\mathcal O_{\rho_\mathrm{upper}}$, the magnitude of the dust energy density is always smaller than that of the scalar field. If above condition is not met, the total energy density  vanishes before the turnaround point occurs, leading to a recollapse of the universe.
Before moving on to next example,  we want to emphasize that  the above results are robust to changes in the value of $\mathcal O_{\varphi_i}$,  namely,  a different initial value of the scalar field will certainly change the pre-inflationary and inflationary e-foldings when the dust field is negligible. However, as the initial dust density increases, the dust field will affect the number of  e-foldings in a qualitatively same way as depicted in Fig. \ref{f4} for any fixed initial value of the scalar field as long as the inflaton is initially released from the left wing of the potential with a positive velocity. In the extreme case, when the inflaton field starts with a large negative value, the inflationary phase would disappear even without the reference clocks. On the other hand, a large positive value of the scalar field can lead to a huge number of e-folding. The effects of the reference clocks would be the same as depicted in Fig. \ref{f4} for a positive initial velocity while the negative initial velocity would have an opposite effect on the e-foldings.

\begin{figure}
{
\includegraphics[width=8.3cm]{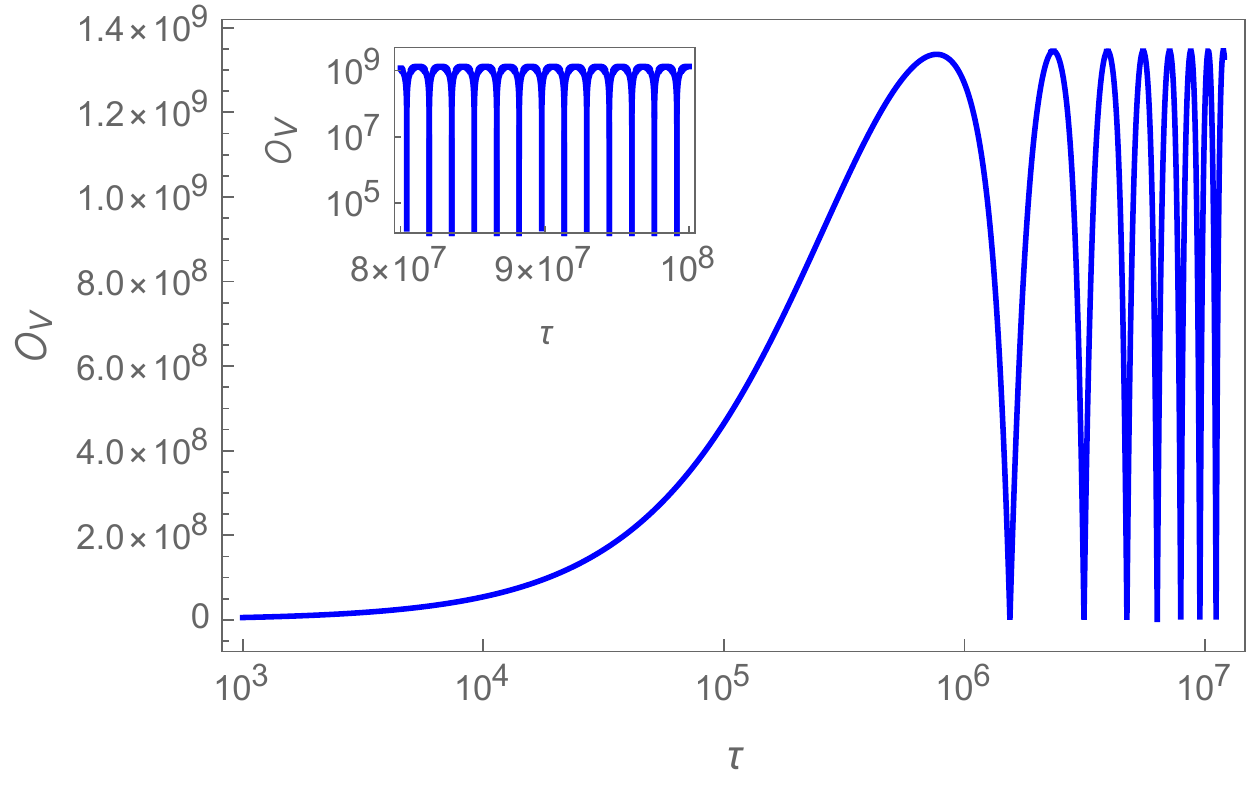}
\includegraphics[width=8cm]{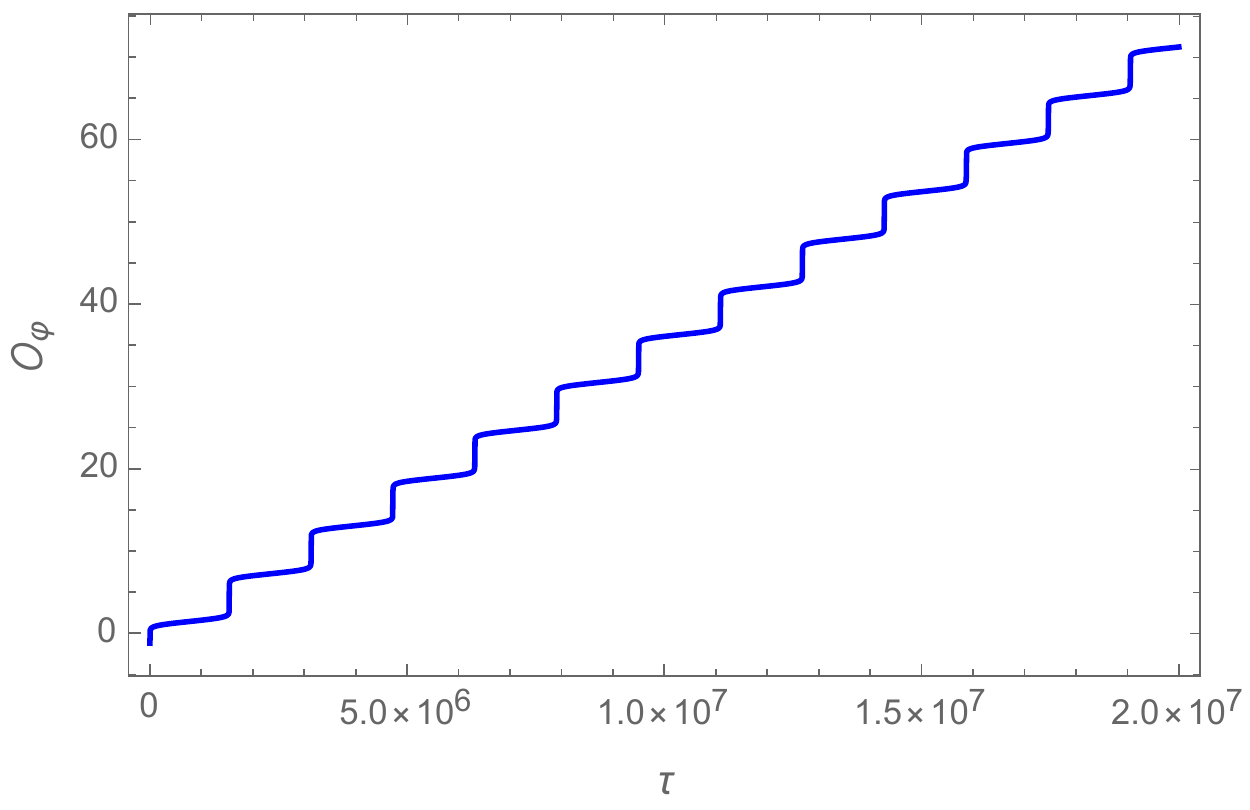}
}
\caption{In this figure,  the initial conditions are chosen as in (\ref{initial3}) for a large negative dust energy density which results in a cyclic universe. The Dirac observable of the volume of the universe increases very slightly in consecutive cycles. The cyclic phase continues for a long time as shown in the inset figure. The value of the Starobinsky field increases steadily in a step-like phenomena.}
\label{f5}
\end{figure}

From our simulations, we find that with the initial scalar field set to $\mathcal O_{\varphi_i}=-1.45$, the upper bound on the magnitude of the initial negative dust energy density turns out to be approximately $\mathcal O_{\rho_\mathrm{upper}}=5.35\times10^{-7}$. When $|\mathcal O_{\rho^\mathrm{dust}_i}|>\mathcal O_{\rho_\mathrm{upper}}$,  the universe undergoes a recollapse and a cyclic dynamics in LQC will appear. This interesting case is not satisfied by ideal dust clocks because a relatively large magnitude of  energy densities changes the universe with only one bounce into the one with many bounces. One such case  occurs with the following initial conditions, chosen again with ${\cal O}_{b_i} =\pi/2\lambda$ and ${O}_{V_i} = 1000$,
\bq
\lb{initial3}
\mathcal O_{\varphi_i}=-1.45 ,\quad \quad \mathcal O_{\rho^\mathrm{dust}_i}=-5.35\times10^{-7},
\eq
which also set $\mathcal O_{\pi_{\varphi_i}}=900$. Results are presented in Fig. \ref{f5} in which the evolution of the Dirac observable of the volume  and the scalar field is depicted. It can be seen from the figure that the universe experiences cycles of alternating expanding and contracting phases. In the process, the volume at successive recollapse points increases very slightly,  showing a hysteresis-like phenomenon which has been discussed recently in a closed universe in LQC \cite{hysteresis}. Meanwhile, the scalar field undergoes a step-like increase at each bounce which is dominated by the kinetic energy of the scalar field. From our numerical solutions, we find that the cyclic universe without the inflationary phase continues until $t=10^{9}$. It is possible that after many bounces and recollapses, the inflation can still takes place at a time when the dust energy density becomes less than the energy density of the scalar field in the expanding phase. However, due to the flatness of the Starobinsky potential on its right wing, the volume of the universe only changes slightly in each cycle, it possibly takes  very long time before the inflation finally sets in. This cyclic behavior of the universe caused by the negative dust energy density in the loop quantized model is in striking contrast with the behavior of the universe in the classical theory. As discussed in \cite{ghls2020}, in the classical FLRW universe, when the magnitude of the negative dust energy density is large enough to cancel the energy density of the scalar field, the universe recollapses and  heads towards the big crunch singularity. However, we find in our case that this singularity is avoided because of quantum gravitational effects and the big crunch singularity is avoided, resulting in a cyclic evolution.
 \begin{figure}
{
\includegraphics[width=8.2cm]{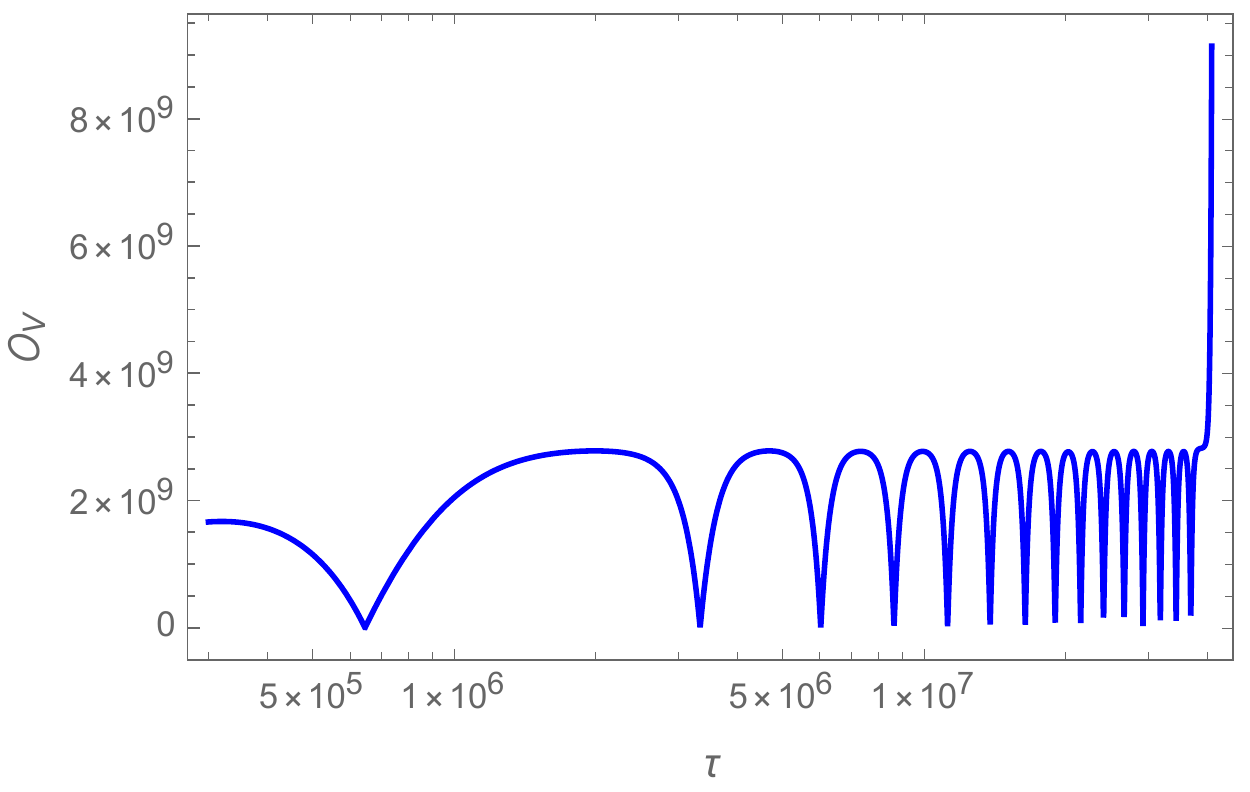}
\includegraphics[width=8cm]{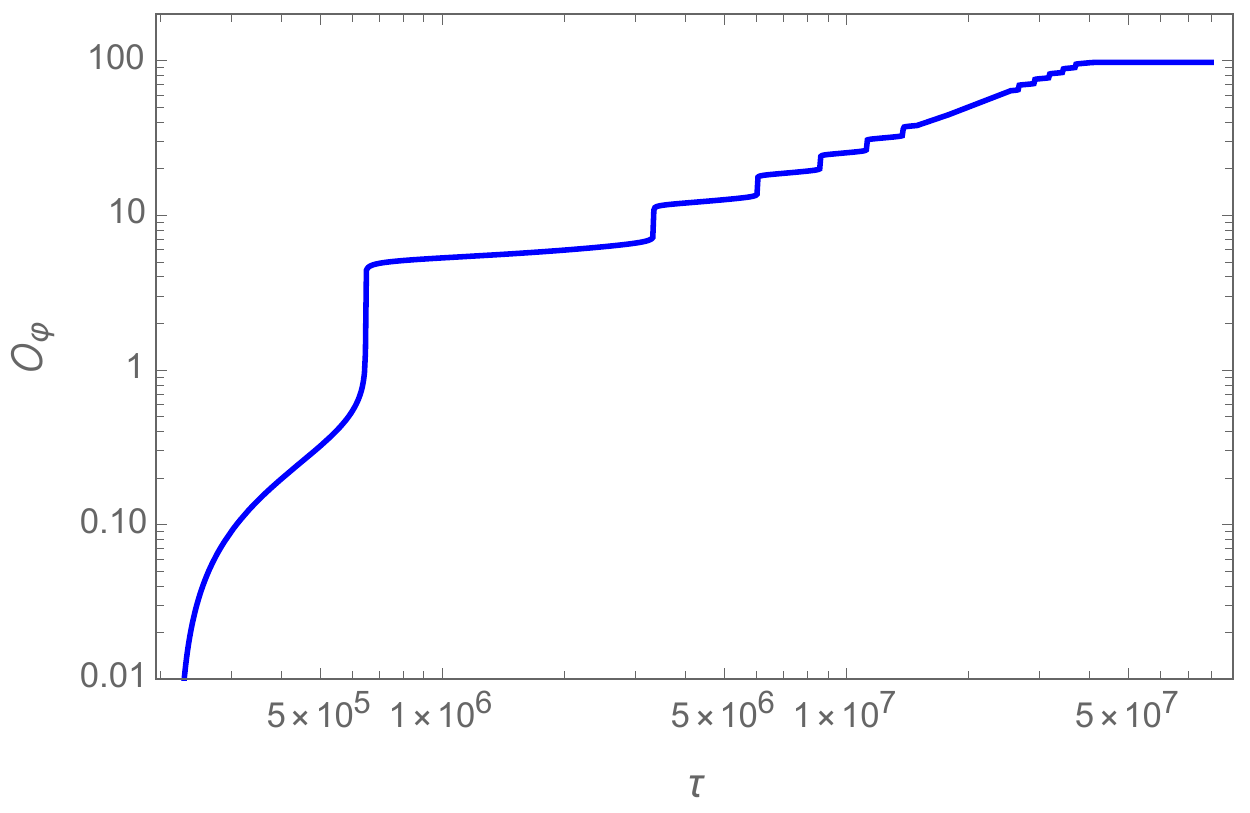}
}
\caption{This figure shows the characteristic evolution of the Dirac observable of the volume and the scalar field when inflation takes place after a few cycles of the expansion and contraction for the Starobinsky potential. The initial conditions are chosen as in (\ref{initial5}) with a negative dust energy density.}
\label{f6}
\end{figure}
While above initial conditions do not easily lead to inflation in short time, it is not difficult to find initial conditions where inflation occurs after some cycles. An example is for the initial conditions:
\bq
\lb{initial5}
\mathcal O_{\varphi_i}=-2.52296,\quad \quad \mathcal O_{\rho^\mathrm{dust}_i}=-10^{-6},
\eq
from which one can find $\mathcal O_{\pi_{\varphi_i}}=905$ using ${\cal O}_{b_i} =\pi/2\lambda$ and ${O}_{V_i} = 1000$. 
The results are shown in Fig. \ref{f6}. The left panel of Fig. \ref{f6} clearly shows that the largest value for the Dirac observable of the volume at different collapse points are changing slightly in the forward evolution. Meanwhile, the scalar field initially  moves steadily in one direction. However, when the inflation takes place, the inflaton reaches its turnaround point and the value of the scalar field starts to decrease. Afterwards, as the volume of the universe experiences an exponential expansion, the effects of the dust field becomes negligible as compared with the scalar field. From our simulations, we find that in order for the universe to transit from the cyclic phase to the inflationary phase, the slope of the scalar potential plays an important role. Unlike the Hubble friction which also assumes oscillating behavior in the cyclic phase, the potential related term in the Klein-Gordon equation (\ref{kg}) always acts as a  frictional force when the scalar field steadily moves in the direction with increasing $\mathcal O_\varphi$.  The hysteresis-like phenomena noted in effective dynamics in LQC in which the volume at the recollapse point grows in successive cycles  \cite{hysteresis} is found explicitly for  this example.

\subsection{Klein-Gordon scalar field clocks}

We now consider the case when a massless Klein-Gordon scalar field plays the role of a clock. For the standard kinetic energy, these clocks have positive energy density. In order to compare with the dust clocks which have positive energy density, we consider representative initial conditions in this case (at the bounce) at ${\cal O}_V = 10^3$ as
\bq
\lb{initial4}
\mathcal O_{\varphi_i}=-1.45,\quad \quad \mathcal O_{\rho_{\chi_i}}=10^{-8},
\eq
here $ \mathcal O_{\rho_{\chi_i}}$ stands for the initial energy density of the massless Klein-Gordon scalar reference field which for comparison with solutions for dust clocks is set equal to the initial dust energy density in (\ref{initial1}). With the initial values given in (\ref{initial4}), one finds $\mathcal O_{\pi_{\varphi_i}}=\pm 900$. Similar to the cases in the dust models, we choose positive  $\mathcal O_{\pi_{\varphi_i}}$ in our numerical analysis. Since the Dirac observable of the lapse is  negative-valued in the Klein-Gordon scalar field model, we compensate this by choosing negative values for  $\tau_\chi$  to study the evolution. As a result, when setting the bounce at $\tau_\chi=0$, inflation takes place in the regime of negative values of $\tau_\chi$.\footnote{Note that in comparison with the unreduced LQC \cite{aps2006c} where Dirac quantization is applied, the Schr\"odinger equation involves a derivative with respect to the scalar field that is chosen as a clock. Because the Dirac observables are a power series in terms of $(\tau_\chi-\chi)$ a derivative of the observable with respect to $\tau_\chi$ and with respect to $\chi$ exactly differ by a minus sign which relates the $\tau_\chi$ used here to the physical time used in \cite{aps2006c}.}
\begin{figure}
{
\includegraphics[width=8.cm]{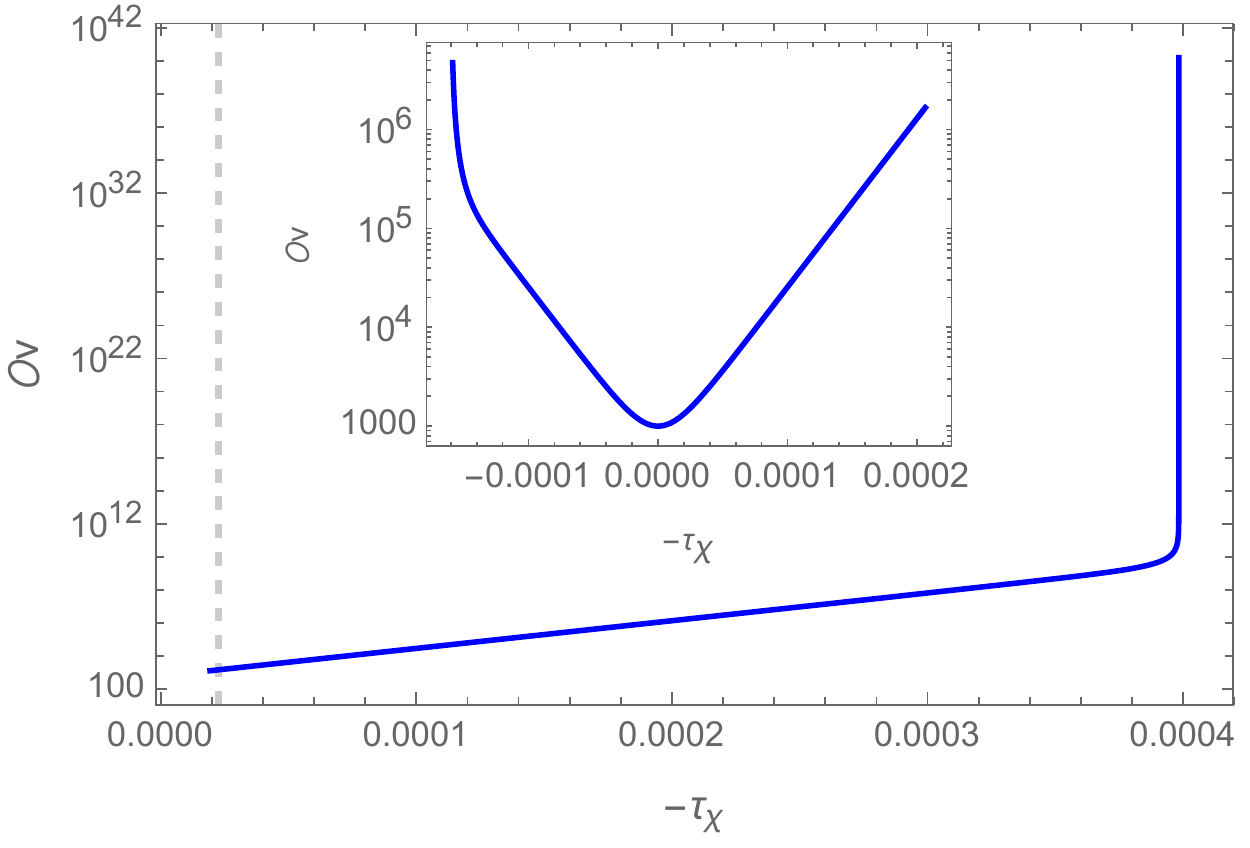}
\raisebox{-0.23cm}{\includegraphics[width=8cm]{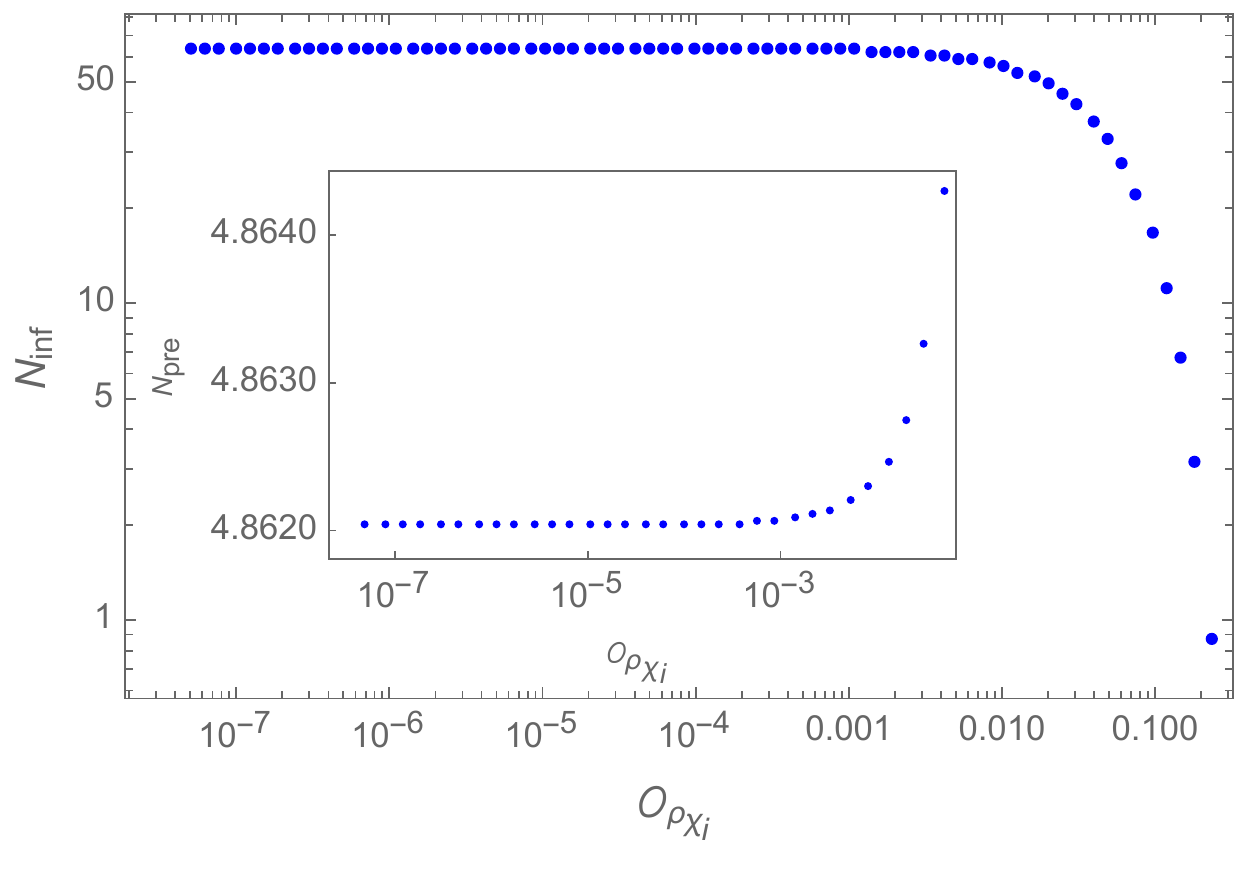}}
}
\caption{In this figure, the initial conditions are chosen at the bounce with the initial values set by (\ref{initial4}) for the Klein-Gordon scalar clock. In  the  left  panel,  we  show  the  continuous  evolution  of  the  Dirac observable of the volume until the onset of inflation. The gray dashed vertical line marks the end of the superinflationary phase at $-\tau_\chi=2.24\times10^{-5}$. The behavior of the the  Dirac observable of the  volume across the bounce is depicted in the inset plot. In the right panel,  we show the impact of the initial energy density of the Klein-Gordon scalar reference field on the pre-inflationary (the inset plot) and the inflationary e-foldings with $\mathcal O_{\varphi_i}=-1.45$.}
\label{f7}
\end{figure}
In Fig. \ref{f7}, we show the evolution of the Dirac observable of the volume and effects of the initial energy density of the Klein-Gordon scalar field  on the background dynamics. In the figure, $\tau_\chi=0$ corresponds to the bounce point, and the super-inflationary phase defined by the region when ${\cal O}_{\rho = \rho_{\mathrm{max}}} \geq {\cal O}_\rho \geq {\cal O}_{\rho =  \rho_{\mathrm{max}}/2}$ ends at $\tau_\chi=-2.24\times10^{-5}$, yielding a total of $0.12$ super-inflationary e-foldings. 
This phase then leads after a short time to slow roll inflation starting  at $\tau_\chi=-3.81\times10^{-4}$  and ending at $\tau_\chi=-3.98\times10^{-4}$, yielding a total of $63.8$ inflationary e-foldings. Note that  due to a faster decay of the energy density of the Klein-Gordon scalar field, the effects of the Klein-Gordon scalar reference field on the background dynamics are quantitatively different from the dust reference field.  To be specific,  starting with the same initial energy densities of the clock fields, the inflationary e-foldings in the two models turn out to be different. As discussed in section \ref{sec:numeric}.A, for the initial conditions (\ref{initial1}), the inflationary e-folding with the dust clocks is $63.1$, while the inflationary e-folding with the Klein-Gordon scalar clocks is $63.8$ which is closer to the inflationary e-foldings, equal to $=63.9$, in the standard LQC effective theory without any reference clocks \cite{zhutao}.  This indicates that  with the same initial energy densities, the Klein-Gordon scalar reference field has less impact on the inflationary e-foldings than the dust reference field. Of course, our numerical results also show that the effects of both clocks can be tuned as negligible as possible if the initial energy density of the clocks are set to small values.

In the right panel of Fig. \ref{f7}, we show the Klein-Gordon scalar reference field  affects the background dynamics in a qualitatively similar way as the dust reference field with a positive energy density.  For a particular  $\mathcal O_{\varphi_i}$ in the parameter space, there also exists an  upper bound on the initial energy density of the Klein-Gordon scalar field  for inflation to occur which turns out to be  0.238 in the current case and is  significantly larger than the upper bound in the dust model (which is $5.35\times10^{-7}$) with the same  $\mathcal O_{\varphi_i}$. Meanwhile, the affect of the Klein-Gordon scalar reference field  on the e-foldings of  the  pre-inflationary and the inflationary regimes is also quantitatively different from the dust reference field with the same initial conditions. For the dust model, as shown in Fig. \ref{f4}, when the initial dust energy density lies between $(10^{-7}, 10^{-4})$, we find a rapid decrease (increase) in the number of the inflationary (pre-inflationary) e-foldings while in the Klein-Gordon scalar field model, the decrease (increase) in the number of the inflationary (pre-inflationary) e-foldings becomes significant only when the initial energy density of the reference field grows above  $0.01$. As a result, the Klein-Gordon scalar reference field serves as a good clock in a larger part of the parameter space  than the dust reference field. 
%%%%%%%%%%%%
%%%%%%%%%%%%
\section{Conclusions}
\label{sec:conclusion}
\renewcommand{\theequation}{5.\arabic{equation}}\setcounter{equation}{0}
The goal of this paper is to understand the loop quantization of  the spatially flat, homogeneous and isotropic FLRW universe sourced with a Starobinsky inflationary potential using  reduced phase space quantization. We used the relational formalism with  Brown-Kucha\v{r}, Gaussian dust and Klein-Gordon scalar reference fields which act as clocks in the quantization procedure. We obtained the physical Hamiltonians in the reduced mini-superspace in terms of Dirac observables associated with four degrees of freedom which are the Dirac observables of the connection and the conjugate triad as well as  the inflaton and its conjugate momentum. We applied the improved dynamics or the $\bar \mu$ scheme in LQC to loop quantize the geometric degrees of freedom. The physical Hilbert space is spanned by the quantum states depending on the volume of the universe,  the value of the inflaton field and the clock time. 
Using properties of the gravitational part of the Hamiltonian and the evolution operator considered generally in LQC \cite{Kaminski:2008td}, the physical Hamiltonian in all three cases turns out to be self-adjoint.
We constructed the quantum theory in terms of operators corresponding to the Dirac observables and obtained the quantum dynamics in terms of their corresponding physical Hamiltonians implemented on the physical Hilbert space.  We showed that the resulting quantum gravitational Schr\"odinger equations are non-singular  quantum difference equations with uniform steps in volume. We investigated some of the physical implications on bounce and inflationary dynamics in the effective spacetime description with the above clocks.   For the reason that we can use the usual LQC representation for the physical Hilbert space in all three models the physical Hilbert space is still non-separable. However, as in (Dirac quantized) LQC, solutions of the quantum gravitational Schr\"odinger equation are expected to obey super-selection sectors and in this case if one restricts to the space of solutions of the quantum dynamics we expect that  the space of solutions becomes a separable Hilbert space. 
Our analysis provides a novel way to bypass various problems with quantization of inflationary potential in Dirac quantization with inflaton as a clock. Unlike in such an approach where the Hamiltonian is time-dependent, the inner product is not conserved and no global clock exists, in our approach one obtains a time-independent Hamiltonian, a conserved inner product and a good global clock, essential for unitarity, is available. Our analysis provides a path to repeat a similar exercise in Dirac quantization by introducing in addition to the inflaton Klein-Gordon or dust reference fields.  The methods used in our analysis can be applied for any potential, including those for alternatives to inflation,  and different quantum cosmological models. To understand aspects of new physics in this model, we used effective dynamics for our two fluid models to find numerical solutions. Effective dynamics has so far in LQC  turned out to be an excellent approximation to the quantum difference evolution equation \cite{aps2006c,dgs2014,djms2017,ps2018}.\footnote{Recently, attempts have been made to obtain effective dynamics in LQC from reduced phase space quantization of LQG using path-integral formulation with dust reference fields \cite{hl2019II}, but such studies do not yet include a potential.} We found the effective Hamiltonians in the reduced mini-superspace and derived the Hamilton's equations and the  modified Friedmann equations for the dust and Klein-Gordon reference field models. As the Dirac observable of a function of the elementary phase space variables is just given by the function of the corresponding Dirac observables, the observable map is in this sense function preserving. Therefore, the modified Friedmann equations in the dust models take formally the same form as their counterpart in the standard LQC in cosmic time while the modified Friedmann equation in the Klein-Gordon scalar field model contains on its right hand side an additional multiplication factor proportional to the square of the Dirac observable of the lapse which is expected since the evolution is formulated with respect to the scalar field time and not the cosmic time. These modified equations effectively describe a dynamical system with two fluids coupled individually to the background spacetime. Besides, these two fluids, i.e. the inflaton field and the clock field, satisfy their individual continuity equations and have no direct couplings between each other.  From the modified Friedmann equations, we found that in the spatially flat FLRW universe of the reduced phase space, the big bang singularity is replaced with a quantum bounce which takes place at the maximum energy density $\rho_\mathrm{max} \approx 0.41 \rho_{\mathrm{Pl}}$ set by the quantum geometry. The bounce density turns out to be the same for different clocks studied in this manuscript.\footnote{Unlike Wheeler-DeWitt theory where the choice of clock has been argued to make a difference in singularity resolution \cite{Gielen:2020abd}, our analysis shows that the same conclusion does not extend to LQC for the clocks we investigated.}

For the numerical analysis, the  initial conditions were chosen at the bounce in a two-dimensional parameter space consisting of the scalar field $\mathcal O_\varphi$ and the energy density of the reference field. The initial conditions for the inflaton were given when the inflaton field was initially  rolling down in the left-wing of the potential. For the positive (negative) dust energy, the presence of the dust field  increases (decreases) the Hubble friction, and thus  decreases (increases) the duration of the inflationary phase for above initial conditions. The dust energy density can also slightly change the duration of the preinflationary phase in the bouncing regime.
We found that when the magnitude of the negative dust energy density exceeds $\mathcal O_{\rho_\mathrm{upper}}$, the total energy density becomes zero before the inflation sets in, causing a recollapse. The  big crunch singularity of classical theory is resolved due to quantum geometric effects  and the universe  undergoes cycles of contracting and expanding phases. The volume of the universe grows slightly in  successive cycles, leading to a change in the dust energy density between any two neighboring cycles reflecting a mild hysteresis-like phenomena  \cite{hysteresis}. With the Starobinsky potential, while some initial conditions do not yield inflation even after long time, we showed the existence of initial conditions which do lead to inflation after various cycles of contraction and expansion\footnote{Similar results have been recently obtained for spatially closed model in  LQC \cite{lss}.}. In all the cases we find non-singular dynamics with dust clocks, in contrast to the classical theory where for the same inflationary potential the universe encounters the big bang singularity in the past, and a big crunch in future for a sufficiently large negative dust energy density \cite{ghls2020}. The effects of the Klein-Gordon scalar field clock turn out to be similar to the dust clock with a positive energy density but with some differences. First, the inflationary phase lasts for a very short period in the time measured by the Klein-Gordon scalar field clock as compared with the dust clock. Secondly, since the energy density of the Klein-Gordon scalar field clock decays much faster than the dust clock, the former has less impact on the duration of inflation  than the latter.  Starting from the same initial energy density of the clock fields, the inflationary e-foldings in the Klein-Gordon scalar field model is larger than in the case of dust reference fields. In general, the Klein-Gordon scalar reference field serves as a good clock in a larger subspace of the whole parameter space than the dust reference field with the positive energy density. We also find that with the same initial conditions, the e-foldings from the bounce to the moment when the pivot mode exits the horizon during inflation are slightly larger for the dust reference fields. This implies the observable window of the primordial power spectrum for the quantum gravitational effects are also changed when different physical clocks are used.   
Note that even though, the Gaussian and Brown-Kucha\v r dust models have exactly the same background dynamics in the spatially flat FLRW universe, from the analysis in \cite{ghls2020}, a difference in the scalar power spectrum is still expected as the equations of motion for the linear perturbations are different in the two dust models. It will be interesting to generalize the latter analysis to LQC and explore the consequences of choosing different dust and scalar clocks on the primordial power spectrum in detail. While above results indicate that starting from the same initial conditions for clock densities, different reference fields can leave tiny but distinct imprints on inflationary dynamics, one can of course choose different initial conditions for different clocks to get same phenomenological effects. In this sense, the multiple choice problem of time, which leaves traces in phenomenology, is linked with  the problem of initial conditions.  Finally, one can always choose initial conditions for the clock densities such that their contribution to observational quantities is less than the experimental error. A more detailed understanding of these models can be obtained by investigating the physical solutions,  and quantum dynamics directly. Compared to the current models in LQC the already existing numerical techniques need to be generalized to the two-fluid case. Due to the presence of a reference field independent of the potential, this is computationally more involved than numerical simulations done so far in isotropic (Dirac quantized) LQC \cite{aps2006c,dgs2014}. The situation is closer to anisotropic models which have more degrees of freedom. Given that the latter have been successfully analyzed in detail using super-computing methods \cite{djms2017,ps2018}, the numerical infrastructure can be adapted to study quantum dynamics in reduced phase space quantization. Apart from the question of singularity resolution at the level of the physical Hilbert space, this step is important since it will allow to test the validity of effective techniques for two-fluid models.
Further, it opens the possibility to  better compare to for instance the model in \cite{aps2006a,aps2006b,aps2006c} at the quantum level, always taking into account that we compare models with different degrees of freedom in the physical Hilbert space. This for instance also involves a better understanding of the possibility to switch  clocks at the quantum level in these models, see for instance \cite{Hoehn:2018aqt,*Hoehn:2018whn} for an analysis in Wheeler-DeWitt cosmology. The notion of quantum reduction maps that have been presented in \cite{Hoehn:2019owq} and used to compare Dirac and reduced quantisation are an interesting approach in this context. 
A long term future project will be, once the quantum background dynamics is well understood, to consider linear perturbations which opens a window to further analyze the different features of the dust and Klein-Gordon scalar fields models. For instance the fact that the physical Hamiltonians in the two dust models coincide is only a special property of the FLRW spacetime and no longer holds if perturbations are taken into account. 
An analysis on above lines would provide a consistent and complete quantum gravitational treatment of inflationary spacetimes where the role of clocks in the background dynamics as well as in perturbations, and important issues such as the physical Hilbert space and the conserved inner product in presence of an  inflationary potential, and resulting impacts on phenomenology will be settled. 

\begin{spacing}{.5}
\vspace{-0.2cm}
\section*{Acknowledgements} 
\vspace{-0.3cm}
This work is  supported by the DFG-NSF grants PHY-1912274 and 425333893 and NSF grant PHY-1454832. K.G. thanks the LSU gravity group for their kind hospitality during a visit where part of this work was done.
\end{spacing}

\begin{spacing}{.5}

\end{spacing}
\end{document}